\documentclass[a4paper,11pt]{article}
\usepackage{jheppub} % for details on the use of the package, please see the JINST-author-manual
\usepackage{lineno}
\usepackage{slashed}
\usepackage{comment}
\def \nn {\nonumber}
\def \be  {\begin{equation}}
\def \ee  {\end{equation}}
\def \bea  {\begin{eqnarray}}
\def \eea  {\end{eqnarray}}

\title{\boldmath Locally infrared finite two-loop amplitudes for multi Higgs production in association with one jet}

\author{Aniruddha Venkata}
\affiliation{Bethe Center for Theoretical Physics,\\  Universit$\ddot{\text{a}}$t Bonn, D-53115, \\Germany}
\preprint{BONN-TH-2026-21}
\emailAdd{avenkata@uni-bonn.de}

\abstract{We study the two-loop amplitudes for multi-Higgs production through a massive top-quark loop in association with one jet. We consider the quark-antiquark channel with a final-state gluon, $q\bar q\to g+nH$ although our construction applies to crossed channel as well. We utilize (non-local in momentum space) Ward identities to factorize longitudinal and gauge degrees of freedom through exact identities between Feynman graphs without reference to an infrared region. In the genuine two loop topologies, we replace the internal gluons by their physical degrees of freedom and show that the resulting amplitude is free of collinear singularities. The only remaining infrared divergence is a soft singularity that can be identified with a change of reference gauge vector. We construct explicit soft counterterms for these singularities, obtaining a locally infrared-finite representation. We verify that the factorized longitudinal degrees of freedom reproduce the remaining infrared poles of the original Feynman graphs. Our formalism requires no counter-terms for collinear finiteness or for manifesting Ward identities in various regions since we perform a decomposition of covariant gauge Feynman graphs. We expect its extension to more complicated processes to eliminate all collinear singularities--particularly in the initial state.}

\begin{document}
\maketitle
\flushbottom

\section{Introduction}
\label{sec:intro}

Two-loop amplitudes with final-state jets are among the missing ingredients needed for precision Higgs studies at the Large Hadron Collider (LHC) \cite{Huss:2025nlt}. In this paper we study the two-loop amplitudes for the process
\bea
q(p_1)+\bar{q}(p_2)\rightarrow g(p_3) + h(q_1) \ldots h(q_m),
\eea
and for its crossed channel, $q+g \to q+mH$. To carry out the two-loop integral numerically in four dimensions, one first has to rewrite the integrand in a form that is infrared (IR) finite.
 
Loop amplitudes in massless gauge theory are studied using an infrared regulator.  A common choice is dimensional regularization in $4+2\epsilon$ dimensions. Infrared divergences of the amplitude appear as poles in $\epsilon$ when using such a regulator. While final-state singularities cancel after summing over all final states (including unresolved parton phase space), initial-state singularities require collinear subtractions in the cross section \cite{Collins:1989gx}. Simple two-loop integrals with a small number of mass scales can be carried out analytically in terms of known functions using dimensional regularization, more complicated integrals with multiple scales remain a bottle neck at two-loops and beyond \cite{Caola:2022ayt}. 

An alternative is to remove the IR divergences before integration, so that the remaining loop integrand can be integrated numerically in exactly $d=4$. Related numerical and local approaches to loop integrals have been developed in Refs.~\cite{Soper:1998ye,Nagy:2003qn,Nagy:2006xy,Gong:2008ww,Binoth:2000ps,Anastasiou:2007qb}.

One way to deal with final-state singularities is the use of time-ordered perturbation theory \cite{Sterman:2023xdj} and the related family of loop-tree duality methods, wherein final-state cancellations can be made manifestly local in momentum space using the unitarity algebra~\cite{Catani:2008xa,Bierenbaum:2010cy,Runkel:2019yrs,Capatti:2019ypt,Aguilera-Verdugo:2020set,Capatti:2020xjf,Capatti:2022tit}.

For real radiation, a class of covariant local subtraction algorithms handles both initial- and final-state singularities in cross sections~\cite{Frixione:1995ms,Catani:1996vz,GehrmannDeRidder:2005cm,Anastasiou:2003gr,Catani:2007vq,Czakon:2010td}.

Focusing on amplitude-level methods, particularly relevant for fully differential predictions with electroweak bosons and jets in the final-state, one would like a locally finite amplitude integrand when the number of scales becomes large \cite{Ma:2019hjq}. One systematic approach to constructing such integrands in gauge theories is the local finiteness program of Anastasiou and Sterman. The authors identify pinch surfaces using the Landau equations~\cite{Landau:1959fi}, which have a simple interpretation in terms of soft lines and jets~\cite{Coleman:1965xm,Sterman:1978bi,Sen:1982bt,Sterman:1993hfp}, and construct local counterterms near each pinch surface. The remainder is finite in four dimensions, while the counterterms are simple enough to integrate analytically. The idea was first demonstrated for massless $2\to2$ master integrals~\cite{Anastasiou:2018rib}, then for two-loop QED amplitudes with several off-shell photons~\cite{Anastasiou:2020sdt}, and subsequently for two-loop QCD amplitudes for $q\bar q\to V_1\dots V_n$~\cite{Anastasiou:2022eym}. It has since been extended to gluon-initiated electroweak and multi-Higgs production through massive quark loops~\cite{Anastasiou:2024xvk} and to real-virtual NNLO corrections~\cite{Anastasiou:2026rvl}. In the QED and QCD constructions, local Ward identities are needed to line up the integrands of different diagrams. The loop-momentum shifts required for this are compensated by explicit ``shift counterterms'', and the subtractions are constructed to be local in loop momentum near each pinch surface. A main drawback of this method is the need for increasingly complicated shift counterterms and, at higher loops, the treatment of jet-soft interactions \cite{Haindl:2025jte}.

In this paper, we study a different construction for the next step in this program: two-loop amplitudes for multi-Higgs production through a massive top-quark loop in association with one jet. We consider the quark-gluon channel with a final-state quark, $qg\to q+mH$, and the crossed quark-antiquark channel with a final-state gluon, $q\bar q\to g+mH$. Our starting point is the G-K decomposition, used in the spirit of the Grammer--Yennie method in QED~\cite{Grammer:1973db}, and applied through exact identities between Feynman graphs as suggested in Ref.~\cite{Venkata:2025lgx}. Using Ward identities between Feynman graphs, without making any reference to an infrared region, we separate the longitudinal and gauge degrees of freedom of the internal gluons from the rest. We then isolate the contribution involving only the physical degrees of freedom of internal gluons. We show that this contribution is free of collinear singularities. The only remaining IR divergence is an overall soft singularity, which we identify with a change of the reference gauge vector, and we construct explicit soft counterterms for it. This gives a locally IR-finite representation of the physical contribution. As a check, we verify that the factorized longitudinal degrees of freedom reproduce the remaining IR poles of the original Feynman graphs.

The main difference from the Anastasiou--Sterman construction is that our method decomposes the amplitude using exact identities that are non-local in momentum space. As a result, we require no shift counterterms. We use the analysis of individual IR regions only to establish that the modified integrand, obtained from our physical-gauge Feynman rules, is finite in collinear limits, without having to construct counterterms from the limits themselves. We expect this method to carry over to more complicated processes and to eliminate all collinear singularities there.

The paper is organized as follows. In Sec.~\ref{sec:kinematics}, we explain the kinematics, introduce a convenient graph classification, and describe some useful graphical tools. In Sec.~\ref{sec:transient}, we use our modified Feynman rules to eliminate ``transient singularities''---singularities that appear in individual graphs but vanish after summing over a gauge-invariant subset. In Sec.~\ref{sec:IR}, we consider the remaining genuinely IR-divergent graphs, which are the main focus of this paper. We show that the two-sided physical-propagator GG terms are collinear finite and construct their soft counterterms. We then show that the GK and KK terms factorize into products of lower-loop amplitudes and can be evaluated consistently. In App.~\ref{sec:check}, we verify that the longitudinal contributions reproduce the IR poles of the original graphs. Finally, Sec.~\ref{sec:conclusions} contains our conclusions.

\section{Kinematics and conventions}
\label{sec:kinematics}
In this section, we will describe the process of interest, list families of graphs that contribute to it, and finally end with a definition of the scalar-scalar-gluon decomposition as well as the G-K decomposition which will be our main tools in the sections that follow. 

In this paper, we will study the process
\bea
q(p_1,s_1,i_1)+\bar{q}(p_2,s_2,i_2)\rightarrow g(p_3,\epsilon_3,x) + h(q_1) \ldots h(q_m),
\eea
where $s_1,s_2$ are the spins of the quark and anti-quark and $i_1,i_2$ are the color indices of the quark and anti quark respectively. We will denote by $\epsilon_3^{\mu}(p_3;p_1)$, the polarization vector of the gluon which satisfies $\epsilon_3^{\mu}(p_3;p_1) \cdot p_3=\epsilon_3^{\mu}(p_3;p_1)\cdot p_1=0$. We will also consistently use $x$ for the (adjoint) color index of the external gluon. All partons will be assumed to be massless, $p_i^2=0$.
\begin{figure}
    \centering
    \includegraphics[width=0.7\linewidth]{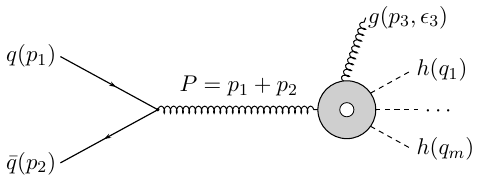}
    \caption{Set of Feynman graphs that describe the process of interest at one loop. Throughout, we will utilize $\mathcal{H}$ to represent a fermion loop, which consists of all topologically distinct permutations of attachments on the top loop. }
    \label{fig:process}
\end{figure}
The one-loop Feynman graphs that contribute to the process are displayed in Fig.\ \ref{fig:process}. It will be useful to represent the fermion loop by the number of distinct electroweak and gluon attachments. If we denote by $\mathcal{H}_{\mu_1\ldots \mu_n}^{a_1\ldots a_n}[n,m](p_1\ldots p_n,q_1\ldots q_m)$ the single fermion loop and by $\mathcal{M}^{(L)}(p_1,p_2,p_3,\{q_i\};s_1,s_2,\epsilon_3,i_1,i_2,x)$ the $L$ loop amplitude, which we will sometimes shorten to $\mathcal{M}^{(L)}$, the one loop amplitude is given by
\bea
\mathcal{M}^{(1)}=\epsilon_3^{\mu_2}(p_3;p_1)\mathcal{H}_{\mu_1\mu_2}^{a,x}[2,m](P, -p_3,q_1\ldots q_m) \frac{-i\delta^{\mu_1}_{\nu}}{P^2+i0}\bar{v}(p_2,s_2)-ig_sT^{a}\gamma^{\nu}u(p_1,s_1).
\eea
Throughout this paper, the fermion-loop function $\mathcal{H}$ includes all coupling factors, fermion propagators, and sums over distinct permutations of gluon and Higgs attachments within the massive top-quark loop.  The fermion loop is defined with incoming gluons and outgoing electroweak bosons. We will also find it useful to use the shorthand $P=p_1+p_2$. 
As an example, $\mathcal{H}[2,1]$ is simply the familiar one-loop top-quark contribution to Higgs production through gluon fusion, with the two gluons allowed to be off shell. Explicitly,
\bea
\mathcal{H}_{\mu\nu}^{ab}[2,1](k_1,k_2;q)
&=&
-\int \frac{d^d\ell}{(2\pi)^d}\,
\mathrm{Tr}_{c,D}\bigg[
(-ig_s T^a\gamma_\mu)\,
S_t(\ell+k_1)\,
(-ig_s T^b\gamma_\nu)\,
S_t(\ell+q)\,
(-iy_t)\,
S_t(\ell)
\bigg]
\nonumber\\
&&
+\big((a,\mu,k_1)\leftrightarrow(b,\nu,k_2)\big),
\eea
where $q=k_1+k_2$, $y_t$, and $S_t(k)=i(\slashed{k}+m_t)/(k^2-m_t^2+i0)$ is the massive top-quark propagator. The trace runs over Dirac and color indices and the minus sign comes from the fermion loop.

\begin{center}
    \begin{figure}
        \centering
        \includegraphics[width=0.8\linewidth]{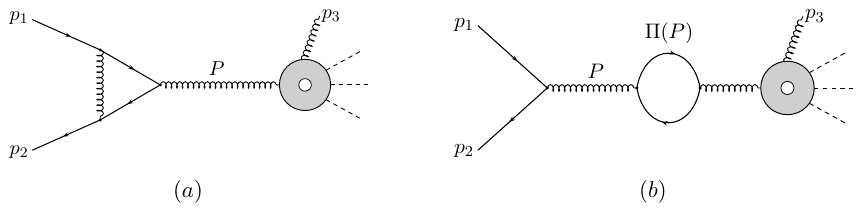}
        \caption{Two examples of topologically factorized graphs. (a) is a vertex correction of the current which factorizes from the fermion loop on account of the  intermediate gluon propagator being a one particle reducible cut. (b) represents the self-energy of the intermediate gluon which also factorizes from the fermion loop.  }
        \label{fig:fact}
    \end{figure}
\end{center}

Let us now classify all graphs that appear in the two-loop amplitude. First, we decompose the two-loop amplitude into three pieces:
\begin{equation}
{\mathcal{M}^{(2)}=\mathcal{M}^{(2)}_{\mathrm{IR}}
+\mathcal{M}^{(2)}_{\overline{\mathrm{IR}}}
+\mathcal{M}^{(2)}_{\mathrm{fact}}.}
\label{eq:split}
\end{equation}
The first two terms contain connected two-loop graphs.
We distinguish between graphs in
$\mathcal{M}^{(2)}_{\mathrm{IR}}$, which require an explicit
treatment of their infrared singularities, and graphs in
$\mathcal{M}^{(2)}_{\overline{\mathrm{IR}}}$, whose infrared
singularities cancel after summing over the appropriate
families of graphs using Ward identities.
The graphs in $\mathcal{M}^{(2)}_{\mathrm{fact}}$
are topologically factorized. 
They satisfy
\bea
\mathcal{M}^{(2)}_{\mathrm{fact}}
=
\sum_{G_{\mathrm{fact}}=G_1\times G_2}
\left[
\int\frac{d^dk}{(2\pi)^d}G_1(k)
\right]
\left[
\int\frac{d^d\ell}{(2\pi)^d}G_2(\ell)
\right].
\eea
Two representative examples of topologically factorized graphs are drawn in Fig.\ \ref{fig:fact}. 
After introducing the scalar-scalar-gluon decomposition, we will be able to classify the remaining graphs into $\mathcal{M}^{(2)}_{\overline{\mathrm{IR}}}$, $\mathcal{M}^{(2)}_{{\mathrm{IR}}}$ which we turn to now. 
\subsection{Scalar-Scalar-Gluon decomposition}
Let us now discuss the first of two tools that we will make use of extensively. It will be convenient to decompose the triple gluon vertex into its three constituent terms. Such a decomposition has already appeared in the literature in \cite{Anastasiou:2024xvk} and was called the scalar-scalar gluon decomposition there. In conventions where all three momenta are incoming into a three gluon vertex,
the momenta satisfy
\begin{equation}
k_1+k_2+k_3=0.
\end{equation}
The full three gluon vertex is given by
\begin{equation}
 V_{\alpha\beta\gamma}^{abc}(k_1,k_2,k_3)
 =-g_sf^{abc}\Big[g_{\alpha\beta}(k_1-k_2)_\gamma
+g_{\beta\gamma}(k_2-k_3)_\alpha
+g_{\gamma\alpha}(k_3-k_1)_\beta\Big].\label{eq:triple-gluon}
\end{equation}
Each term in this vertex will now be represented by a distinct graph using the decomposition
\begin{center}
\includegraphics[]{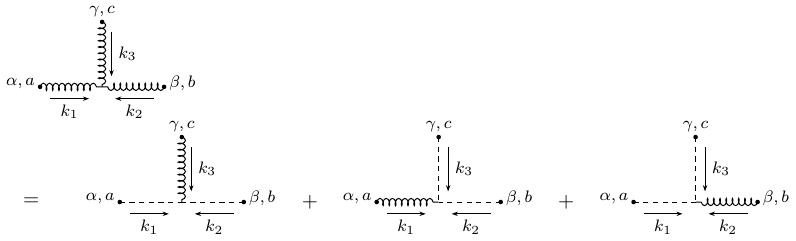}%
\end{center}
Here, each graph is given by only one term in Eq.\ (\ref{eq:triple-gluon}). Explicitly, we have
\begin{center}
\includegraphics[width=.30\linewidth]{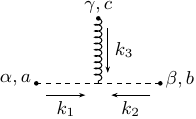}
\end{center}
\begin{equation}
\bigl[V^{abc}\bigr]_{\alpha\beta\gamma}^{(12)}
=-g_sf^{abc}g_{\alpha\beta}(k_1-k_2)_\gamma.
\end{equation}
Throughout this paper, we will use the notation $[V]$ to remind the reader that the vertex is not the full three gluon vertex, but rather a specific piece picked out through this decomposition. The specific term  picked out in Eq.\ \ref{eq:triple-gluon} will be made clear using a graph. 
With this graph decomposition on hand, we are now ready to split all topologies further into $\mathcal{M}^{(2)}_{\overline{\mathrm{IR}}}$,$\mathcal{M}^{(2)}_{{\mathrm{IR}}}$. A further subclassification within $\mathcal{M}^{(2)}_{\overline{\mathrm{IR}}}$  distinguishes between the pinched graphs which need a graph sum to manifest finiteness from the unpinched graphs which are automatically infrared finite. 
\bea
\mathcal{M}^{(2)}_{\overline{\mathrm{IR}}}=\mathcal{M}^{(2)}_{\overline{\mathrm{IR}},\text{pinched}}+\mathcal{M}^{(2)}_{\overline{\mathrm{IR}},\text{unpinched}}\label{eq:IRbar-catalogue-split}
\eea
\subsection{List of non-factorized two-loop graphs}
\label{sec:graph-catalogue}

Using the scalar-scalar-gluon decomposition, we can now classify
all the remaining non-factorized two-loop graphs into the two families
$\mathcal{M}^{(2)}_{\mathrm{IR}}$ and
$\mathcal{M}^{(2)}_{\overline{\mathrm{IR}}}$ introduced in
Eq.~\eqref{eq:split}. Throughout this catalogue, $\mathcal H$
represents the complete massive fermion loop, including all distinct
permutations of gluon and Higgs attachments.

\begin{figure}[t]
  \centering
  \begingroup
  \newcommand{\IRpanel}[2]{%
    \begin{minipage}[c]{#1}
      \centering
      \includegraphics[
        width=.98\linewidth,
        height=3.0cm,
        keepaspectratio
      ]{figures/IR_#2.pdf}\\[1.5mm]
      {\small \(\mathrm{(#2)}\)}
    \end{minipage}%
  }

  \IRpanel{.315\linewidth}{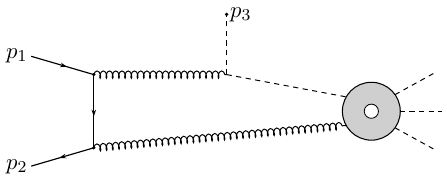}\hfill
  \IRpanel{.315\linewidth}{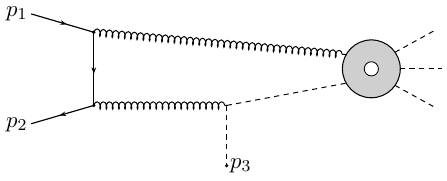}\hfill
  \IRpanel{.315\linewidth}{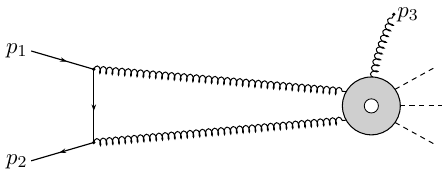}

  \par\vspace{3mm}

  \IRpanel{.46\linewidth}{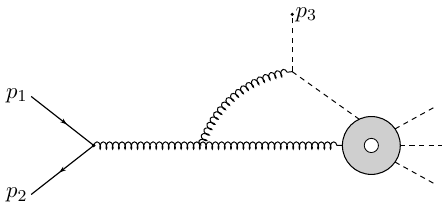}\hspace{.035\linewidth}
  \IRpanel{.46\linewidth}{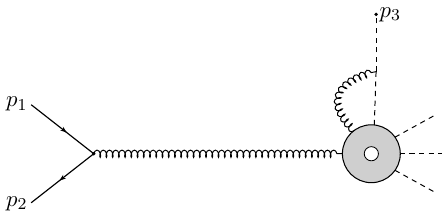}

  \endgroup

  \caption{The five graph topologies contributing to
    $\mathcal{M}^{(2)}_{\mathrm{IR}}$.
    Dashed lines denote scalar legs of the scalar-scalar-gluon decomposition.}
  \label{fig:IR-catalogue}
\end{figure}

The first family is shown in Fig.~\ref{fig:IR-catalogue}.
These five graphs have genuine soft or collinear singularities, whose
leading contributions must be isolated by factorization. We return
to them in Sec.~\ref{sec:IR}.

The remaining graphs belong to
$\mathcal{M}^{(2)}_{\overline{\mathrm{IR}}}$.
It is useful to separate them further according to whether their
massless propagators admit infrared pinch surfaces as in Eq.\ \ref{eq:IRbar-catalogue-split}

\begin{figure}[h]
    \centering
    \includegraphics[width=.98\linewidth]
    {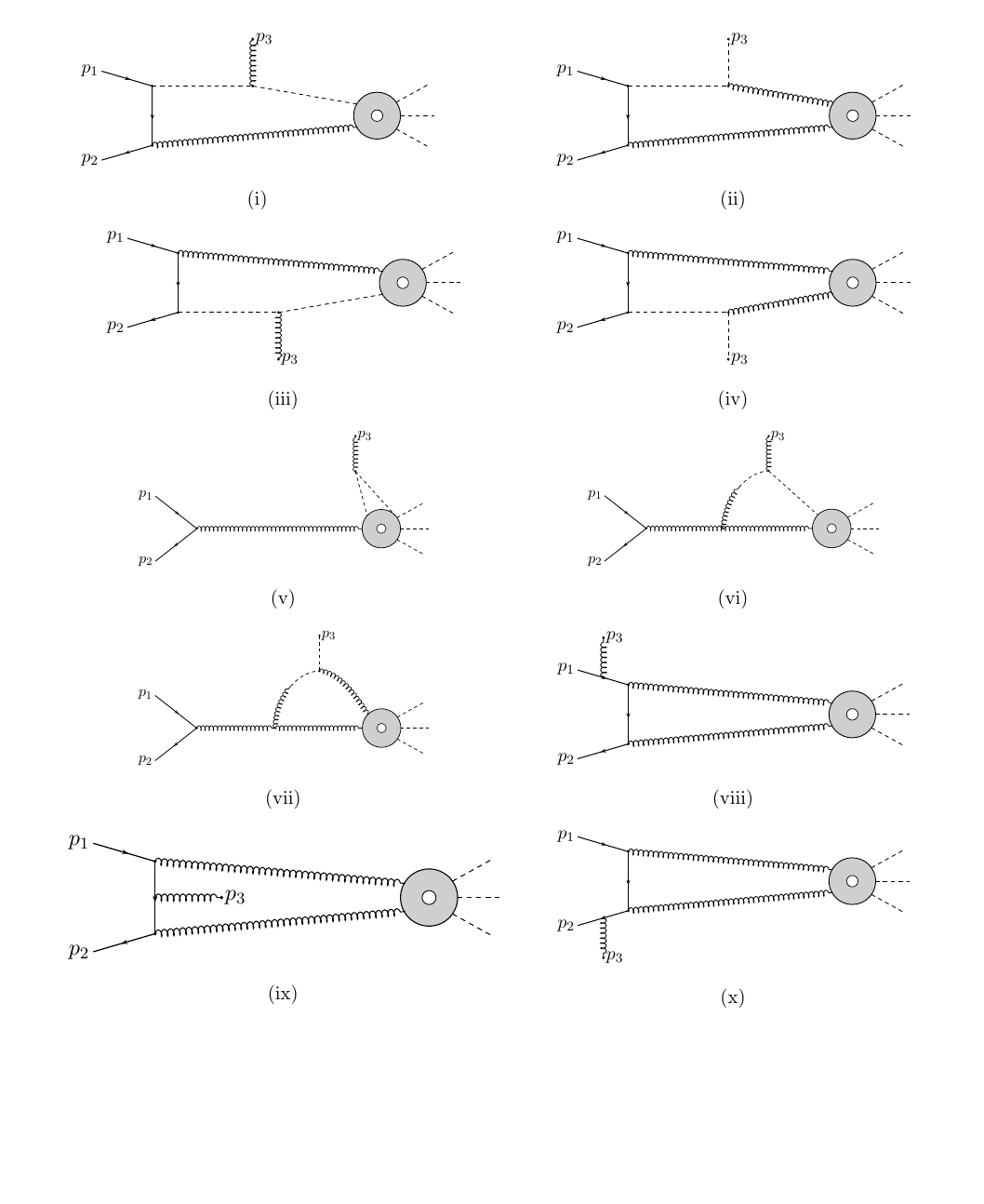}
    \caption{The ten pinched graphs contributing to
    $\mathcal{M}^{(2)}_{\overline{\mathrm{IR}}}$,
    labeled (i)--(x). }
    \label{fig:IRbar-pinched-catalogue}
\end{figure}

The ten pinched graphs are displayed in
Fig.~\ref{fig:IRbar-pinched-catalogue}.
Although their denominators admit soft or collinear pinches, their
infrared singularities disappear after the appropriate Ward-identity
rearrangements and graph sums. The corresponding analysis is given
in Sec.~\ref{sec:transient}.

\begin{figure}[h]
    \centering
    \includegraphics[width=.98\linewidth]
    {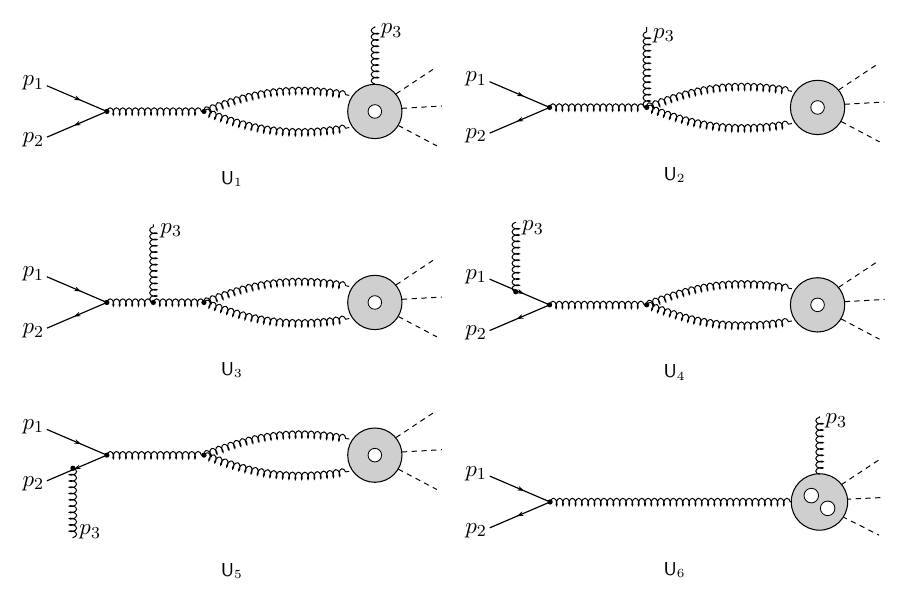}
    \caption{The unpinched contribution to
    $\mathcal{M}^{(2)}_{\overline{\mathrm{IR}}}$.}
    \label{fig:IRbar-unpinched-catalogue}
\end{figure}

Finally, Fig.~\ref{fig:IRbar-unpinched-catalogue} contains
the unpinched graphs, which require no infrared rearrangement. The last graph, labeled $(\alpha)$,
denotes the compact two-loop hard-loop contribution.

\subsection{G-K decomposition}

We will also make extensive use of the G-K
decomposition of the gluon propagator.
For a gluon carrying momentum $k$, we define
\bea
\delta^\mu_{\ \nu}
&=&
G^\mu_{\ \nu}(k;q)
+
K^\mu_{\ \nu}(k;q),
\\
K^\mu_{\ \nu}(k;q)\label{eq:GK}
&=&
\frac{(2q+k)^\mu k_\nu}
{k^2+2q\cdot k+i0},
\\
G^\mu_{\ \nu}(k;q)
&=&
\delta^\mu_{\ \nu}
-
K^\mu_{\ \nu}(k;q).
\eea
Here $q$ is a reference momentum and will be taken to be time-like throughout, satisfying $q^2 >0$. A specific choice that works well is $q=P=p_1+p_2$. With this choice the auxilliary denomiantor is, $(k+P)^2-P^2+i0$. 
The $K$ term is longitudinal with respect to
the gluon momentum $k$ and will be treated
using Ward identities.
Away from the pole of the projector,
we have
\bea
k_\mu G^\mu_{\ \nu}(k;q)&=&0,\label{eq:G-Kpow}
\\
G^\mu_{\ \nu}(k;q)(2q+k)^\nu&=&0. 
\eea
We will sometimes utilize the appropriately raised/lowered $G-K$ tensors, defined by $G_{\mu}^{\;\;\nu}=\eta_{\mu\rho}G^{\rho}_{\;\sigma}\eta^{\sigma\nu}$.
Let us interpret these equations, in a manner useful to the discussion that follows.
The G gluon  supplies power suppression in the soft limit because $G^{
\mu}_{\;\nu}(k;q)(2q+k)^\nu=0$ only if it attaches to a jet with lightlike momentum $q$. However, when it is chosen to be time-like as we do here, gauge freedom (the freedom to shift reference vectors for gluons) ensures the difference between the power suppressed G gluon with reference vector $p$ (jet momentum) and reference vector $q$ (arbitrary reference) is a longitudinal degree of freedom (in the soft limit). 

The first equation in Eq.\ (\ref{eq:G-Kpow}) will find use in collinear limits when $k\parallel p_i$ where $p_i$ is some light-like vector. This identity suggests that when the G gluon attaches to a $p_i$ jet, it is automatically power suppressed.  

Finally, we will frequently apply the G-K decomposition
at both ends of a gluon propagator.
Writing the momentum at the opposite end as
$k_1=-k$, we have
\bea
\frac{-ig_{\mu\nu}}{k^2+i0}
&=&
\frac{\big[G^\alpha_{\ \mu}(k;q)
+K^\alpha_{\ \mu}(k;q)\big]
-ig_{\alpha\beta}
\big[G^\beta_{\ \nu}(k_1;q)
+K^\beta_{\ \nu}(k_1;q)\big]}{k^2+i0}
\nonumber\\
&=&-i
\frac{(GgG)_{\mu\nu}
+(GgK)_{\mu\nu}
+(KgG)_{\mu\nu}
+(KgK)_{\mu\nu}}{k^2+i0}.\label{eq:gluon-prop-GK-2sided}
\eea
The two-sided decomposition separates the
gluon propagator into four contributions.
Terms containing longitudinal polarizations
can be simplified using Ward identities,
while the remaining contributions will be finite in all collinear limits. Their soft limits  will be
treated explicitly with counterterms. 

\begin{figure}[t]
    \centering
    \includegraphics[width=.66\linewidth]
    {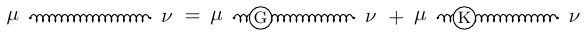}

    \vspace{1em}

    \includegraphics[width=.98\linewidth]
    {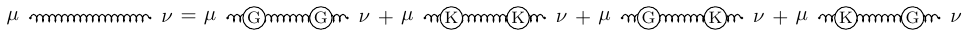}

    \caption{Modified Feynman rules corresponding to
    the one-sided G-K decomposition in
    Eq.~\eqref{eq:GK} (top) and the two-sided
    G-K decomposition in
    Eq.~\eqref{eq:gluon-prop-GK-2sided} (bottom).
    The circles represent the $G$ and $K$ projectors,
    with the propagator denominator and overall
    factor of $-i$ understood.}
    \label{fig:GK-propagators}
\end{figure}

Before ending this section, let us introduce graphical notation for the G-K  when applied to a gluon propagator.
We will use the diagrammatic identities in
Fig.~\ref{fig:GK-propagators} as modified Feynman rules
throughout this paper. These identities are exact and
are in one-to-one correspondence with the one-sided G-K decomposition in
Eq.~\eqref{eq:GK} and the two-sided decomposition in
Eq.~\eqref{eq:gluon-prop-GK-2sided}, respectively.
The circles labeled $G$ and $K$ represent the corresponding
projectors acting on the Lorentz indices of the gluon
propagator.

\section{Transient infrared singularities}
\label{sec:transient}
In this section, we begin our study of the singular graphs with transient singularities. These are all graphs in $\mathcal{M}_{\overline{\text{IR}}, \text{pinched}}^{(2)}$ as it appears in Eq.\ (\ref{eq:IRbar-catalogue-split}) and in Fig.\ \ref{fig:IRbar-pinched-catalogue}. 

\begin{center}
    \begin{figure}[ht]
        \centering
        \includegraphics[width=1.05\linewidth]{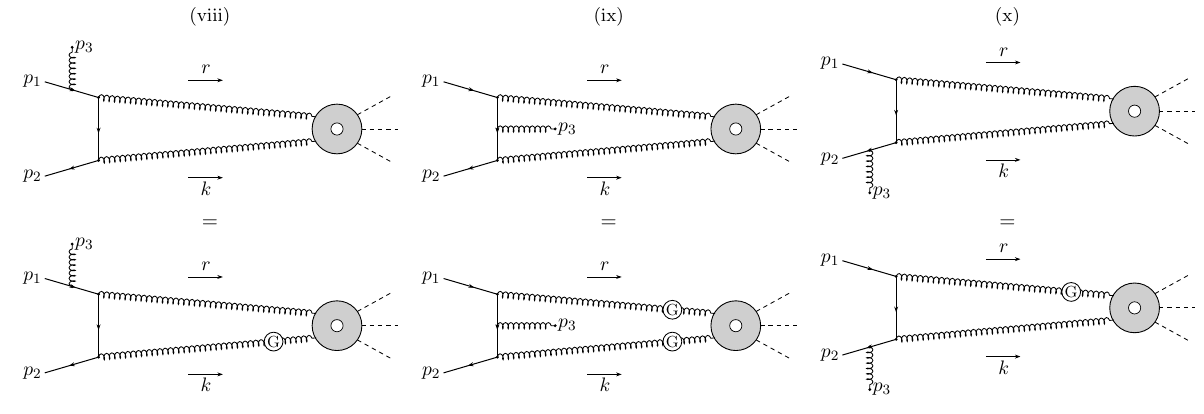}
        \caption{The ward identity based replacement Feynman rules for graph topologies (viii)-(x). The top and bottom panels are exactly equal after hard loop integration and diagram sum.}
        \label{fig:transient-id-1}
    \end{figure}
\end{center}

Let us begin by considering graphs (viii)--(x) of Fig.~\ref{fig:IRbar-pinched-catalogue}.
These graphs have two gluons attached to the massive fermion loop, while
all three gluon vertices on the open light-quark line are ordinary quark--gluon
vertices.  We will label one of the gluons entering the fermion loop with loop momentum $k$ and the other by $r=\sum\limits_{i=1}^{n}q_i-k$. Sometimes, we will use the shorthand $Q^\mu=\sum\limits_{i=1}^{n}q_i$ (to be distinguished from $q$, the reference vector for the G-K decomposition). The strategy here will be to first write their Feynman integrands and then apply the one-sided
$G$--$K$ decomposition. The decomposition will utilize the integrated fermion loop ward identity to make the integrand locally infrared finite. The result of this replacement is shown in Fig.\ \ref{fig:transient-id-1}. 
 We will utilise the concise symbols for vertex factors and propagators
\begin{equation}
 S_0(l)=\frac{i\slashed l}{l^2+i0},\qquad
 D_{\mu\nu}(l)=\frac{-ig_{\mu\nu}}{l^2+i0},\qquad
 \Gamma_a^\mu=-ig_sT^a\gamma^\mu,\qquad
 \Gamma_3=-ig_sT^x\slashed\epsilon_3(p_3;p_1).
 \label{eq:transient-3-rules}
\end{equation}
Recall that $\mathcal H$ includes the permutation sum and the massive
fermion-loop integration. Let us now study the integrands with the propagators of the
two gluons entering $\mathcal H$ explicitly written, 
\begin{align}
 \mathcal I_{\rm viii}(k)
 &={}
 \bar v(p_2)\,\Gamma_a^\mu S_0(k-p_2)
 \Gamma_b^\rho S_0(p_1-p_3)\Gamma_3\,u(p_1)
 D_{\mu}{}^{\nu}(k)D_{\rho}{}^{\sigma}(r)
 \mathcal H^{ab}_{\nu\sigma}[2,m](k,r;\{q_j\}),
 \label{eq:transient-3-viii}\\
 \mathcal I_{\rm ix}(k)
 &={}
 \bar v(p_2)\,\Gamma_a^\mu S_0(k-p_2)
 \Gamma_3 S_0(p_1-r)\Gamma_b^\rho\,u(p_1)
 D_{\mu}{}^{\nu}(k)D_{\rho}{}^{\sigma}(r)
 \mathcal H^{ab}_{\nu\sigma}[2,m](k,r;\{q_j\}),
 \label{eq:transient-3-ix}\\
 \mathcal I_{\rm x}(k)
 &={}
 \bar v(p_2)\,\Gamma_3 S_0(p_3-p_2)
 \Gamma_a^\mu S_0(p_1-r)\Gamma_b^\rho\,u(p_1)
 D_{\mu}{}^{\nu}(k)D_{\rho}{}^{\sigma}(r)
 \mathcal H^{ab}_{\nu\sigma}[2,m](k,r;\{q_j\}).
 \label{eq:transient-3-x}
\end{align}
Here the three expressions correspond, respectively, to emission of $p_3$
next to the $p_1$ endpoint, between the two gluons going into $\mathcal H$,
and next to the $p_2$ endpoint. In Eqs.~\eqref{eq:transient-3-viii}--
\eqref{eq:transient-3-x} the symmetry factors of these graphs have been suppressed since the cancellation only requires the fermion loop ward identity. More precisely, each graph contribution
is
$\int d^dk\,\frac{1}{|G|}\mathcal I_{
m graph}(k)/(2\pi)^d$.

Because the fermion loop has only two external gluons and all other
attachments are color singlets, its integrated Ward identities read
\begin{equation}
 k^\nu\mathcal H^{ab}_{\nu\sigma}[2,m](k,r;\{q_j\})=0,
 \qquad
 r^\sigma\mathcal H^{ab}_{\nu\sigma}[2,m](k,r;\{q_j\})=0.
 \label{eq:transient-3-H-WI}
\end{equation}
We can therefore replace {each} of the two gluon propagators by its
one-sided $G$ term, without changing the result at fixed $k,r$:
\begin{align}
 D_{\mu}{}^\nu(k)\mathcal H_{\nu\sigma}
 &=\frac{-i}{k^2+i0}
 G_{\mu}{}^\nu(k;q)\mathcal H_{\nu\sigma},
 \label{eq:transient-3-Gk}\\
 D_{\rho}{}^\sigma(r)\mathcal H_{\nu\sigma}
 &=\frac{-i}{r^2+i0}
 G_{\rho}{}^\sigma(r;q)\mathcal H_{\nu\sigma}.
 \label{eq:transient-3-Gr}
\end{align}
Here, we have use the fact that the $K$ gluonic  replacement is
\begin{equation}
 \frac{-i}{k^2+i0}K_{\mu}{}^\nu(k;q)\mathcal H_{\nu\sigma}
 =\frac{-i}{k^2+i0}
 \frac{(2q+k)_\mu\,k^\nu\mathcal H_{\nu\sigma}}
 {k^2+2q\cdot k+i0}=0,
 \label{eq:transient-3-Kzero}
\end{equation}
and similarly for $r$. Throughout this section we use the shorthand
\begin{equation}
  D^{G}{}_{\mu}{}^{\nu}(k) \equiv \frac{-i}{k^{2}+i0}\, G_{\mu}{}^{\nu}(k;q),
\end{equation}
which is equal to $D_{\mu}{}^{\nu}(k)$ when contracted with $\mathcal{H}$,
since the $K$ part vanishes by the Ward identity in Eq.\ (\ref{eq:transient-3-H-WI}).   After this replacement, the new integrands are
\begin{align}
 \mathcal I_{\rm viii}^{(G)}(k)
 &={}
 \bar v(p_2)\,\Gamma_a^\mu S_0(k-p_2)
 \Gamma_b^\rho S_0(p_1-p_3)\Gamma_3\,u(p_1)
 D_{\mu}^{(G)}{}^{\nu}(k)D_{\rho}{}^{\sigma}(r)
 \mathcal H^{ab}_{\nu\sigma}[2,m](k,r;\{q_j\}),
 \label{eq:transient-3-viii-G}\\
 \mathcal I_{\rm ix}^{(G)}(k)
 &={}
 \bar v(p_2)\,\Gamma_a^\mu S_0(k-p_2)
 \Gamma_3 S_0(p_1-r)\Gamma_b^\rho\,u(p_1)
 D_{\mu}^{(G)}{}^{\nu}(k)D_{\rho}^{(G)}{}^{\sigma}(r)
 \mathcal H^{ab}_{\nu\sigma}[2,m](k,r;\{q_j\}),
 \label{eq:transient-3-ix-G}\\
 \mathcal I_{\rm x}^{(G)}(k)
 &={}
 \bar v(p_2)\,\Gamma_3 S_0(p_3-p_2)
 \Gamma_a^\mu S_0(p_1-r)\Gamma_b^\rho\,u(p_1)
 D_{\mu}{}^{\nu}(k)D_{\rho}^{(G)}{}^{\sigma}(r)
 \mathcal H^{ab}_{\nu\sigma}[2,m](k,r;\{q_j\}).
 \label{eq:transient-3-x-G}
\end{align}

 Here, we have made an explicit choice in the reference vector that defines the $K$ polarized longitudinal gluon. Specifically,  we choose the same fixed timelike reference vector $P$ in both projectors 
$P^2>0$ for the massive color-singlet final state. Before discussing the behavior of the modified graphs near pinches of the original graph, let us discuss the two limits in which the newly introduced denominator vanishes. First, it vanishes in the soft region where $k\rightarrow 0$. However, in this region, assuming canonical soft scaling ($k^\mu\rightarrow \lambda Q; \lambda\rightarrow 0$), the newly added term  is $\frac{k^{\nu}}{q\cdot k}$ which is $O(1)$. The denominator can also vanish when $k^{\mu}$ is space-like, with a potential pinched subgraph with on-shell  auxiliary denominator and another denominator which is vanishing. For the graphs occurring in this section, such a pinch can be explicitly excluded when $q=P$ for generic kinematic configurations.

Let us now discuss infrared fineness of these graphs in the IR limits where the original graphs were singular. 
Consider the graph (viii) which was singular in the $k\parallel p_2$ limit. The original singularity was logarithmic. Let us check this statement. Decompose \bea k=\alpha_1 p_2+\alpha_{2}\bar{p}_2+k_{\perp},\eea with $\bar{p}_2=(p_2^0,-\vec{p_2})$and observe that \bea\bar{v}(p_2)\Gamma^{\mu}_aS_0(k-p_2)\frac{\eta_{\mu\nu}}{k^2+i0}=\bar{v}(p_2)-igT_a(\alpha-1)\frac{2p_2^{\mu}}{(k-p_2)^2+i0}\frac{\eta_{\mu\nu}}{k^2+i0}+O(k_{\perp})\label{eq:coll-8},\eea up to terms that are power suppressed in the collinear limit. One can check that the other denominators not appearing here are all finite in the collinear limit. As a result, one finds that the assuming collinear pinch scaling $k_{\perp}\rightarrow \sqrt{\lambda} Q $, $\alpha_2 \rightarrow \lambda$, in the uniform $\lambda\rightarrow0$ limit, one finds a logarithmic divergence (both the measure and the denominator in Eq.\ (\ref{eq:coll-8}) vanish uniformly as $\lambda^2$). 
However, the G gluon replacement that we have made in Eqs.\ (\ref{eq:transient-3-Gr}) induces additional power suppression since
\bea
\frac{p_2^{\mu}(G_{\mu\nu})}{k^2+i0}=0+O(k_{\perp}).
\eea
This proves power counting finiteness of (viii) in the $k\parallel p_2$ limit. Let us briefly discuss the same mechanism of power suppression near the pinch in the other two graphs. 
Graph (x) was singular in the $r\parallel p_1$ limit. The singularity is again logarithmic and arises from the propagators $r^2$ and $(p_1-r)^2$. However, the replacement of the gluon propagator by its G part gives
\bea
\frac{p_1^\mu G_{\mu\nu}(r;q)}{r^2+i0}
=\frac{O(r_\perp)}{r^2+i0},
\qquad r\parallel p_1.
\eea
Thus, the original logarithmic singularity is power suppressed, and graph (x) is locally finite.

Graph (ix) has both of the collinear limits discussed above, namely $k\parallel p_2$ and $r\parallel p_1$. Since we have made the G replacement on both gluon propagators, we find
\bea
p_2^\mu G_{\mu\nu}(k;q)=O(k_\perp),
\qquad
p_1^\mu G_{\mu\nu}(r;q)=O(r_\perp).
\eea
The original logarithmic singularities are therefore removed in both limits. There is one additional pinch surface, where the two internal quark propagators become collinear to $p_3$. Writing their momenta as $\ell$ and $\ell+p_3$, the leading numerator vanishes because
\bea
\slashed{\ell}\slashed{\epsilon}_3
(\slashed{\ell}+\slashed p_3)
=O(\ell_\perp),\qquad \ell\parallel p_3,
\eea
where we have used $\epsilon_3\cdot p_3=0$. Thus graph (ix) is locally finite in this limit as well.

\begin{center}
    \begin{figure}[ht]
        \centering
        \includegraphics[width=1.05\linewidth]{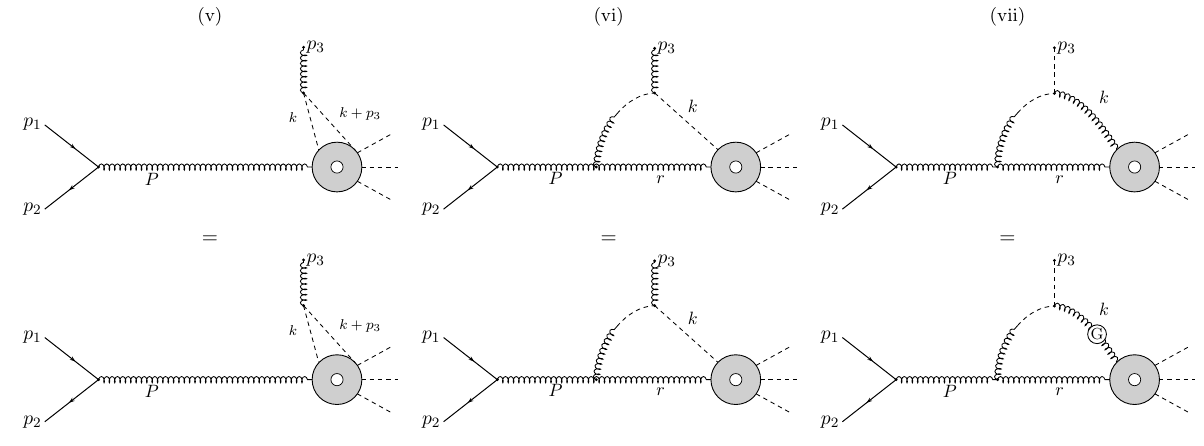}
        \caption{Graphs (v),(vi), and (vii). While the first two are already locally finite, the last is made finite using the G-K decomposition. }
        \label{fig:5-7-rules}
    \end{figure}
\end{center}

We now consider graphs (v)--(vii). In all three cases, the only pinch surface is reached when an internal loop momentum is collinear to $p_3$. We will find that while (v),(vi) are already locally finite, we may once again utilize the one sided G-K decomposition to ensure that (vii) is locally finite. The graphs and the replacement Feynman rules are summarized in Fig.\ \ref{fig:5-7-rules}

The Feynman rules for the graphs in Fig. \ref{fig:5-7-rules} in the notation of Eq.~\eqref{eq:transient-3-rules},  at the level of the three integrands are
\begin{align}
\mathcal I_{\rm v}(k)
={}&\bar v(p_2)\Gamma_a^\alpha u(p_1)
 D_{\alpha}{}^{\mu}(P)
 \mathcal H^{abc}_{\mu\rho\sigma}[3,m](P,k,-k-p_3;\{q_j\})
 \notag\\
&\times D^{\rho\eta}(k)D^{\sigma\theta}(k+p_3)
 [V^{bcx}]_{\eta\theta\lambda}(-k,k+p_3,-p_3)
 \epsilon_3^\lambda,
\label{eq:transient-v-integrand}\\[2pt]
\mathcal I_{\rm vi}(k)
={}&\bar v(p_2)\Gamma_a^\alpha u(p_1)
 D_{\alpha}{}^{\mu}(P)
 V_{\mu\eta\zeta}^{abc}(P,k-Q,-k-p_3)
 \notag\\
&\times D^{\eta\nu}(Q-k)D^{\zeta\beta}(k+p_3)
 [V^{cdx}]_{\beta\gamma\lambda}(k+p_3,-k,-p_3)
 \notag\\
&\times D^{\gamma\sigma}(k)\mathcal H^{bd}_{\nu\sigma}[2,m](Q-k,k;\{q_j\})
 \epsilon_3^\lambda,
\label{eq:transient-vi-integrand}\\[2pt]
\mathcal I_{\rm vii}(k)
={}&\bar v(p_2)\Gamma_a^\alpha u(p_1)
 D_{\alpha}{}^{\mu}(P)
 V_{\mu\eta\zeta}^{abc}(P,k-Q,-k-p_3)
 \notag\\
&\times D^{\eta\nu}(Q-k)D^{\zeta\beta}(k+p_3)
 [V^{cxd}]_{\beta\lambda\gamma}(k+p_3,-p_3,-k)
 \notag\\
&\times D^{\gamma\sigma}(k)\mathcal H^{bd}_{\nu\sigma}[2,m](Q-k,k;\{q_j\})
 \epsilon_3^\lambda.
\label{eq:transient-vii-integrand}
\end{align}
Here we have suppressed the numerical symmetry factors, as in Eqs.~\eqref{eq:transient-3-viii}--\eqref{eq:transient-3-x}.

The scalar-scalar-gluon vertices in (v) and (vi) give, respectively,
\begin{equation}
\begin{aligned}
\epsilon_3^\lambda [V^{bcx}]_{\eta\theta\lambda}(-k,k+p_3,-p_3)
 &=2g_sf^{bcx}g_{\eta\theta}\,\epsilon_3\cdot k,\\
\epsilon_3^\lambda [V^{cdx}]_{\beta\gamma\lambda}(k+p_3,-k,-p_3)
 &=-2g_sf^{cdx}g_{\beta\gamma}\,\epsilon_3\cdot k.
\end{aligned}
\label{eq:transient-v-vi-vertex}
\end{equation}
Since $\epsilon_3\cdot p_3=0$, both numerators are $O(k_{\perp})$ when $k\parallel p_3$. Thus (v) and (vi) are locally finite without any propagator modification.

For (vii), the two scalar legs instead carry $p_3$ and $k+p_3$, so
\begin{equation}
\epsilon_3^\lambda [V^{cxd}]_{\beta\lambda\gamma}(k+p_3,-p_3,-k)
 =-g_sf^{cxd}\epsilon_{3\beta}(2p_3+k)_\gamma.
\label{eq:transient-vii-vertex}
\end{equation}
The gluon of momentum $k$ enters the complete $\mathcal H[2,m]$, and its Ward identity allows us to replace $D(k)$ by $D^G(k;q)$, exactly as above. Concretely, the new integrand for $\text{(vii)}$ reads
\begin{align}
\mathcal I_{\rm vii}^{(G)}(k)
={}&\bar v(p_2)\Gamma_a^\alpha u(p_1)
 D_{\alpha}{}^{\mu}(P)
 V_{\mu\eta\zeta}^{abc}(P,k-Q,-k-p_3)
 \notag\\
&\times D^{\eta\nu}(Q-k)D^{\zeta\beta}(k+p_3)
 [V^{cxd}]_{\beta\lambda\gamma}(k+p_3,-p_3,-k)
 \notag\\
&\times D^{\gamma\sigma}_G(k)\mathcal H^{bd}_{\nu\sigma}[2,m](Q-k,k;\{q_j\})
 \epsilon_3^\lambda.
\label{eq:transient-vii-integrand-G}
\end{align}
Since $k_\gamma G^\gamma{}_{\sigma}(k;q)=0$,
\begin{equation}
(2p_3+k)_\gamma G^\gamma{}_{\sigma}(k;q)
 =2p_{3\gamma}G^\gamma{}_{\sigma}(k;q)
 =O(k_{\perp}),\qquad k\parallel p_3.
\label{eq:transient-vii-G-power}
\end{equation}
In all three cases, the graphs are power counting finite. 

\begin{center}
    \begin{figure}
        \centering
        \includegraphics[width=1.05\linewidth]{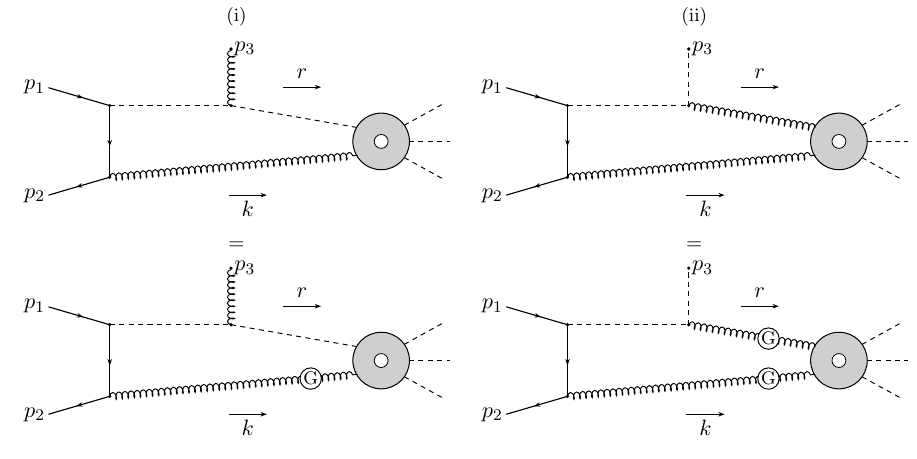}
        \caption{One sided G-K replacement rules for graphs (i),(ii). The K remainder is identically zero as before.}
        \label{fig:1-2rules}
    \end{figure}
\end{center}

Let us now turn to the final class of graphs, (i)-(iv). These have potential singularities in all collinear limits as well as the soft limit. For graphs (i),(ii) we may continue using the one sided G-K decomposition to manifest local finiteness. The result of this is shown in Fig.\ \ref{fig:1-2rules}

We use the momentum routing shown in Fig.~\ref{fig:1-2rules}, where
\bea
P=p_1+p_2=k+r+p_3,\qquad Q=k+r.
\eea
The two gluons entering the massive fermion loop carry momenta $k,r$, and the propagator between the two gluon attachments on the quark line carries momentum $k-p_2$. The integrands after the G replacement are
\bea
\mathcal I_j^{(G)}(k)
={}&
\bar v(p_2)\Gamma_c^\mu S_0(k-p_2)
\Gamma_d^\alpha u(p_1)\,
D^G_{\mu}{}^\nu(k;q)
D_{\alpha}{}^\eta(r+p_3)
\nonumber\\
&\times
\epsilon_3^\lambda
[V^{dex}]^{(j)}_{\eta\beta\lambda}
(r+p_3,-r,-p_3)
D_j^{\beta\rho}(r;q)
\mathcal H_{\nu\rho}^{ce}[2,m](k,r),\label{eq:Greplacement-1,2}
\eea
where
\bea
D_{\mathrm i}(r;q)=D(r),\qquad
D_{\mathrm{ii}}(r;q)=D^G(r;q).
\eea
The Feynman rules give the relations
\bea
\epsilon_3^\lambda
[V^{dex}]^{(\mathrm i)}_{\eta\beta\lambda}
(r+p_3,-r,-p_3)
&=-2g_sf^{dex}g_{\eta\beta}\,
\epsilon_3\cdot r,
\\
\epsilon_3^\lambda
[V^{dex}]^{(\mathrm{ii})}_{\eta\beta\lambda}
(r+p_3,-r,-p_3)
&=g_sf^{dex}\epsilon_{3\eta}
(r+2p_3)_\beta.
\eea

Let us first consider the collinear limits. In the $k\parallel p_2$ limit, both graphs have the same logarithmic singularity arising from the propagators $k^2$ and $(k-p_2)^2$. As before, the G replacement gives
\bea
p_2^\mu G_{\mu\nu}(k;q)=O(k_\perp),
\eea
which makes the singularity locally integrable.

Next, consider $r+p_3\parallel p_1$. In graph (i), the decomposed numerator satisfies
\bea
\epsilon_3\cdot r
=\epsilon_3\cdot(r+p_3),
\eea
which is power suppressed since $\epsilon_3\cdot p_1=0$. In graph (ii), the leading quark numerator in either the soft or collinear to $p_1$ limit is proportional to $p_1^\eta$, whose contraction with $\epsilon_{3\eta}$ vanishes for the same reason.

Finally, in the $r\parallel p_3$ limit, the singular propagators are $r^2$ and $(r+p_3)^2$. The numerator of graph (i) is suppressed by $\epsilon_3\cdot r$, while in graph (ii) we have
\bea
(r+2p_3)^\beta G_{\beta\rho}(r;q)
=2p_3^\beta G_{\beta\rho}(r;q)
=O(r_\perp).
\eea
Thus both graphs are locally finite in all three collinear limits.

For $k\to0$, only $k^2$ and $(k-p_2)^2$ vanish, scaling as $\lambda^2$ and $\lambda$, respectively. Similarly, for $r\to0$, only $r^2$ and $(r+p_3)^2$ vanish, with the same scaling. Both limits are therefore integrable without requiring any additional numerator suppression.

The remaining soft limit is $r+p_3\to0$. Here three denominators vanish: $(r+p_3)^2$, $r^2$ and $(k-p_2)^2$, scaling as $\lambda^2,\lambda,\lambda$. In graph (i), $\epsilon_3\cdot r=\epsilon_3\cdot(r+p_3)$ supplies one additional power of $\lambda$. In graph (ii), the leading quark numerator vanishes because $\epsilon_3\cdot p_1=0$, giving the same suppression. Hence this limit is also integrable.

\begin{center}
    \begin{figure}[ht]
        \centering
        \includegraphics[width=1.05\linewidth]{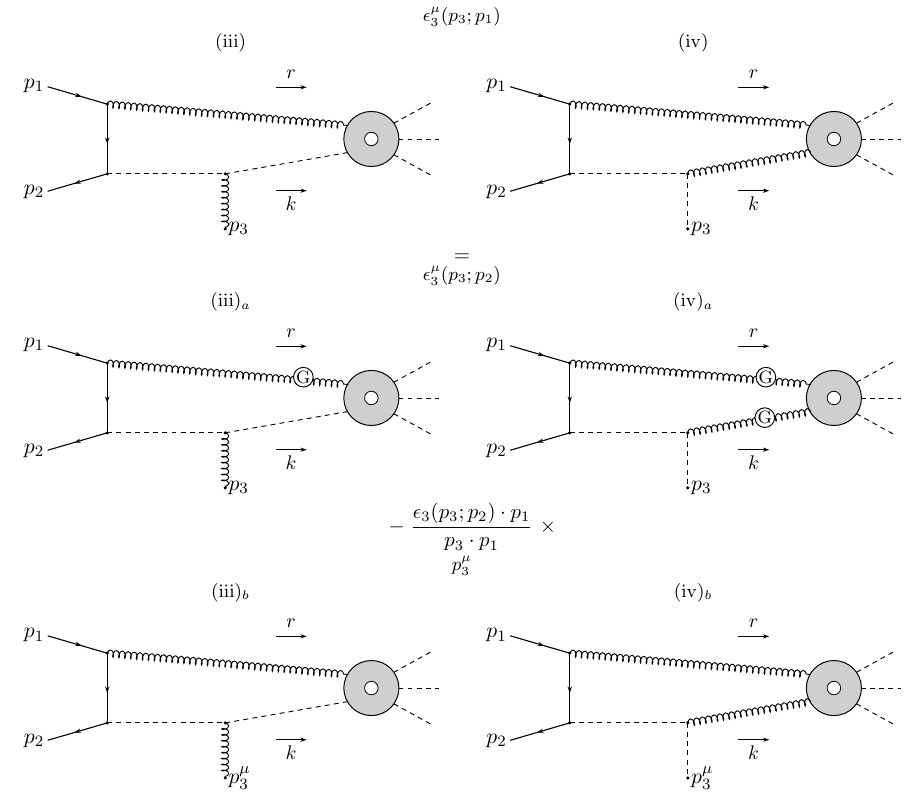}
        \caption{Utilizing the identity of Eq.\ (\ref{eq:change-pol}) in diagram (iii),(iv). In the second panel we have replaced $\epsilon_3(p_3;p_1)$ by $\epsilon_3(p_3;p_2)$, and made the G replacements akin to Fig.\ \ref{fig:1-2rules}. The last panel contains the remainder.}
        \label{fig:3-4rules}
    \end{figure}
\end{center}

We conclude that graphs (i) and (ii) are locally finite in all their infrared limits.
In arguing the finiteness of (i),(ii) after appropriate G gluon replacement, we have utilized the relationship $\epsilon_3\cdot p_1=0$. Although, (iii),(iv) are the class of Feynman graphs generated by switching the roles of $p_1,p_2$, a symmetric argument fails because $\epsilon_3(p_3;p_1)\cdot  p_2\neq 0$. The strategy in this case will be to study (iii),(iv) with a polarization vector that satisfies $\epsilon_3(p_3;p_2)\cdot p_2=0$ and study the gauge vector changing K gluon separately. 
Concretely, we propose to utilize the identity
\bea
\epsilon_3^{\mu}(p_3;p_1)=\epsilon_3^{\mu}(p_3;p_2)- \frac{\epsilon_3(p_3;p_2)\cdot p_1 }{p_3\cdot p_1}p_3^{\mu}.\label{eq:change-pol}
\eea
Replacing the polarization vector $\epsilon_3^{\mu}(p_3;p_1)$ by the right hand side of Eq.\ (\ref{eq:change-pol}) in graphs (iii),(iv), the first term can now be made locally finite in the same way as before. We will focus on properties of the second term in what follows. We have inserted Eq.\ (\ref{eq:change-pol}) and made the same replacements as in Fig.\ \ref{fig:1-2rules}, and the remainder in Fig. \ref{fig:3-4rules}
Following the arguments made previously (iii)$_a$,(iv)$_a$ are locally finite after the relevant G replacements are made. 
As usual the expression for $\mathcal{I}_{\text{(iii),(iv)}}$ in the notation of Eq.\ (\ref{eq:transient-3-rules}) with  $\ell=k+p_3$ and $r=P-\ell$ is
\bea
\mathcal I_{\mathrm{iii}}(k;\epsilon_3)
&=&
-g_s f^{cdx}\,
\bar v(p_2)\Gamma^c_{\rho}
S_0(\ell-p_2)\Gamma^b_{\eta}u(p_1)\,
D^{\rho\mu}(\ell)D^{\eta\tau}(r)
\nonumber\\
&&\times
g_{\mu\nu}(2k+p_3)\cdot\epsilon_3\,
D^{\nu\sigma}(k)\,
\mathcal H^{db}_{\sigma\tau}[2,m](k,r),
\label{eq:iii-integrand}\\
\mathcal I_{\mathrm{iv}}(k;\epsilon_3)
&=&
g_s f^{cdx}\,
\bar v(p_2)\Gamma^c_{\rho}
S_0(\ell-p_2)\Gamma^b_{\eta}u(p_1)\,
D^{\rho\mu}(\ell)D^{\eta\tau}(r)
\nonumber\\
&&\times
\epsilon_{3\mu}(k+2p_3)_\nu\,
D^{\nu\sigma}(k)\,
\mathcal H^{db}_{\sigma\tau}[2,m](k,r).
\label{eq:iv-integrand}
\eea
The splitting equation described in Fig.\ (\ref{fig:3-4rules}) is given by
\bea
\mathcal I_{\mathrm{iii}}(k;\epsilon_3(p_3;p_1))
&=&\mathcal I_{\mathrm{iii},a}(k)
+\mathcal I_{\mathrm{iii},b}(k),
\nonumber\\
\mathcal I_{\mathrm{iii},a}(k)
&=&\mathcal I_{\mathrm{iii}}(k;\epsilon_3(p_3;p_2)),
\nonumber\\
\mathcal I_{\mathrm{iii},b}(k)
&=&-\frac{\epsilon_3(p_3;p_2)\cdot p_1}{p_3\cdot p_1}
\mathcal I_{\mathrm{iii}}(k;p_3),
\label{eq:iii-split}
\eea
and
\bea
\mathcal I_{\mathrm{iv}}(k;\epsilon_3(p_3;p_1))
&=&\mathcal I_{\mathrm{iv},a}(k)
+\mathcal I_{\mathrm{iv},b}(k),
\nonumber\\
\mathcal I_{\mathrm{iv},a}(k)
&=&\mathcal I_{\mathrm{iv}}(k;\epsilon_3(p_3;p_2)),
\nonumber\\
\mathcal I_{\mathrm{iv},b}(k)
&=&-\frac{\epsilon_3(p_3;p_2)\cdot p_1}{p_3\cdot p_1}
\mathcal I_{\mathrm{iv}}(k;p_3).
\label{eq:iv-split}
\eea
Following Eq.\ \ref{eq:Greplacement-1,2}, in $\text{iii}_a,\text{iv}_a $ we replace the $k,r$ gluons by their G counterparts and find that both graphs are power counting finite. 

\begin{figure}
    \centering
    \includegraphics[width=1.05\linewidth]{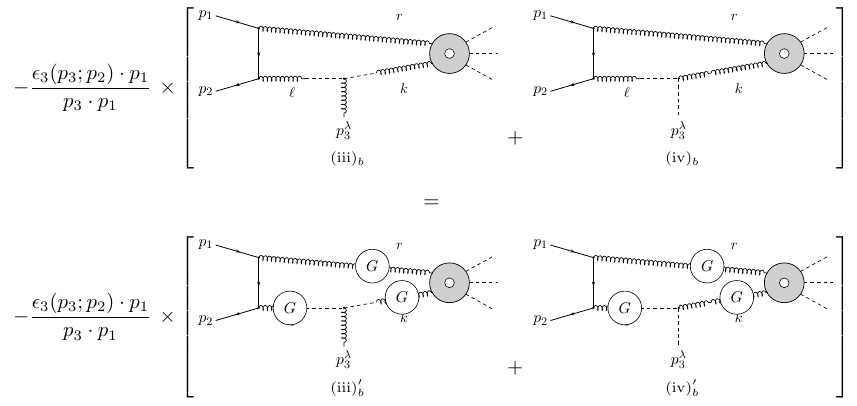}
    \caption{The G gluonic equality in Eqs.\ (\ref{eq:iii-b-G}) and (\ref{eq:iv-b-G})}
    \label{fig:3-4equality}
\end{figure}

Let us turn our attention to (iii)$_b$,(iv)$_b$. Here, the lines labeled by $r,k$ can be replaced by their one sided G components, since as usual, the K components vanish by the fermion loop Ward identity in Eq.\ (\ref{eq:transient-3-H-WI}). However, as we will establish presently, the K components of $l=k+p_3$ also vanishes after summing over $\text{(iii)}_{b},\text{(iv)}_{b}$ and we may replace $l$ propagator by its $G$ component. 
\bea
\mathcal I_{\mathrm{iii},b}^{(G)}(k)
&=&
g_s f^{cdx}
\frac{\epsilon_3(p_3;p_2)\cdot p_1}{p_3\cdot p_1}
\bar v(p_2)\Gamma^c_{\rho}
S_0(\ell-p_2)\Gamma^b_{\eta}u(p_1)
\nonumber\\
&&\times
D^{G,\rho\mu}(\ell;q)
D^{G,\eta\tau}(r;q)
g_{\mu\nu}\,
p_3\cdot(2k+p_3)
D^{G,\nu\sigma}(k;q)
\mathcal H^{db}_{\sigma\tau}[2,m](k,r).
\label{eq:iii-b-G}
\eea

\bea
\mathcal I_{\mathrm{iv},b}^{(G)}(k)
&=&
-g_s f^{cdx}
\frac{\epsilon_3(p_3;p_2)\cdot p_1}{p_3\cdot p_1}
\bar v(p_2)\Gamma^c_{\rho}
S_0(\ell-p_2)\Gamma^b_{\eta}u(p_1)
\nonumber\\
&&\times
D^{G,\rho\mu}(\ell;q)
D^{G,\eta\tau}(r;q)
p_{3\mu}(k+2p_3)_\nu
D^{G,\nu\sigma}(k;q)
\mathcal H^{db}_{\sigma\tau}[2,m](k,r).
\label{eq:iv-b-G}
\eea
This equality is shown in Fig. \ref{fig:3-4equality}
Using the usual power counting relations, one may argue that that the second panel of Fig. \ref{fig:3-4equality} is indeed power counting finite. 

Before concluding this section, let us check that the K gluonic component of $l$ vanishes identically in the polarization changing terms. 
The K component of the $\ell=k+p_3$ propagator contracts the Scalar-Scalar-Gluon vertex with $\ell^\mu$. Summing the two SSG vertices in Eqs.\ (\ref{eq:iii-integrand}) and (\ref{eq:iv-integrand}), and contracting with the longitudinal polarization $p_3^\lambda$, we find
\bea
&&\ell^\mu p_3^\lambda
\left[
-g_s f^{cdx}g_{\mu\nu}(2k+p_3)_\lambda
+g_s f^{cdx}g_{\lambda\mu}(k+2p_3)_\nu
\right]
\nonumber\\
&=&
g_s f^{cdx}\left[
-(\ell^2-k^2)\ell_\nu
+(\ell\cdot p_3)(k+2p_3)_\nu
\right]
\nonumber\\
&=&-g_s f^{cdx}(k\cdot p_3)k_\nu,
\label{eq:iii-iv-longitudinal-WI}
\eea
where we have used $\ell=k+p_3$ and $p_3^2=0$. The remaining longitudinal momentum $k_\nu$ vanishes upon contraction with the hard loop, since
\bea
k_\nu D^{\nu\sigma}(k)
\mathcal H^{db}_{\sigma\tau}[2,m](k,r)
=
\frac{-i}{k^2+i0}\,
k^\sigma\mathcal H^{db}_{\sigma\tau}[2,m](k,r)
=0.
\label{eq:iii-iv-K-vanishing}
\eea
Thus, the K component of the $\ell$ propagator cancels exactly between (iii)$_b$ and (iv)$_b$, and the sum is unchanged upon replacing this propagator by its G component, as shown in Fig.\ \ref{fig:3-4equality}.

To summarize, in this section, we have seen that all graphs with transient infrared divergences can be made locally finite by replacing some propagators by appropriately defined G gluonic counterparts. While the exact replacement needed varies from graph to graph, in each case, we utilize a ward identity that is manifest after the hard loop is integrated over and all permutations within the hard loop are summed over. In the next section, we will turn our attention to the genuinely infrared divergent graphs contained in $\mathcal{M}_{\text{IR}}$. 
\section{Infrared divergent graphs}
\label{sec:IR}
In this section, we study the IR divergent graphs in Fig. \ \ref{fig:IR-catalogue}. In all the graphs of interest, we will apply the  G-K decomposition and find that any graph with at least one instance of a K gluon factorizes into a product of lower loop graphs. The G gluonic graphs will be collinear finite while its soft limit will be handled with explicit counterterms.

\begin{center}
    \begin{figure}
        \centering
        \includegraphics[width=1.05\linewidth]{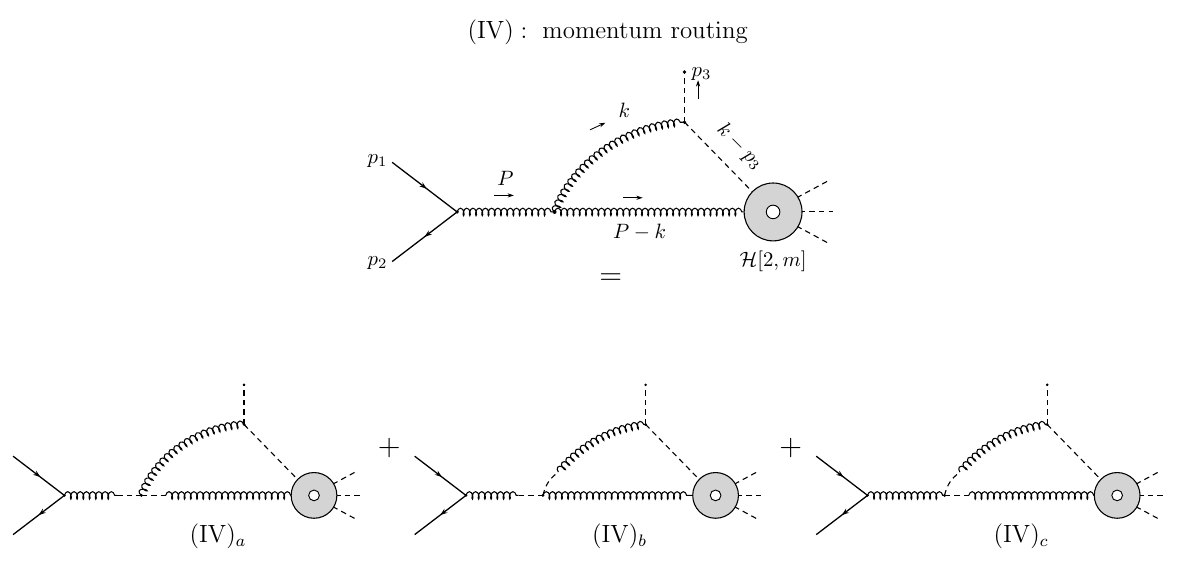}
        \caption{The scalar-scalar-gluon decomposition of graph (IV).}
        \label{fig:IV-decomp}
    \end{figure}
\end{center}

We begin with graph (IV), decomposing its three-gluon vertex into three terms,
\bea
\mathcal I_{\mathrm{IV}}(k)
&=&
\mathcal I_{\mathrm{IV},a}(k)
+\mathcal I_{\mathrm{IV},b}(k)
+\mathcal I_{\mathrm{IV},c}(k).
\label{eq:IV-split}
\eea
The decomposition is shown in Fig.\ \ref{fig:IV-decomp}
We will show that the last two terms are locally infrared finite.

\begin{center}
    \begin{figure}
        \centering
        \includegraphics[width=1.05\linewidth]{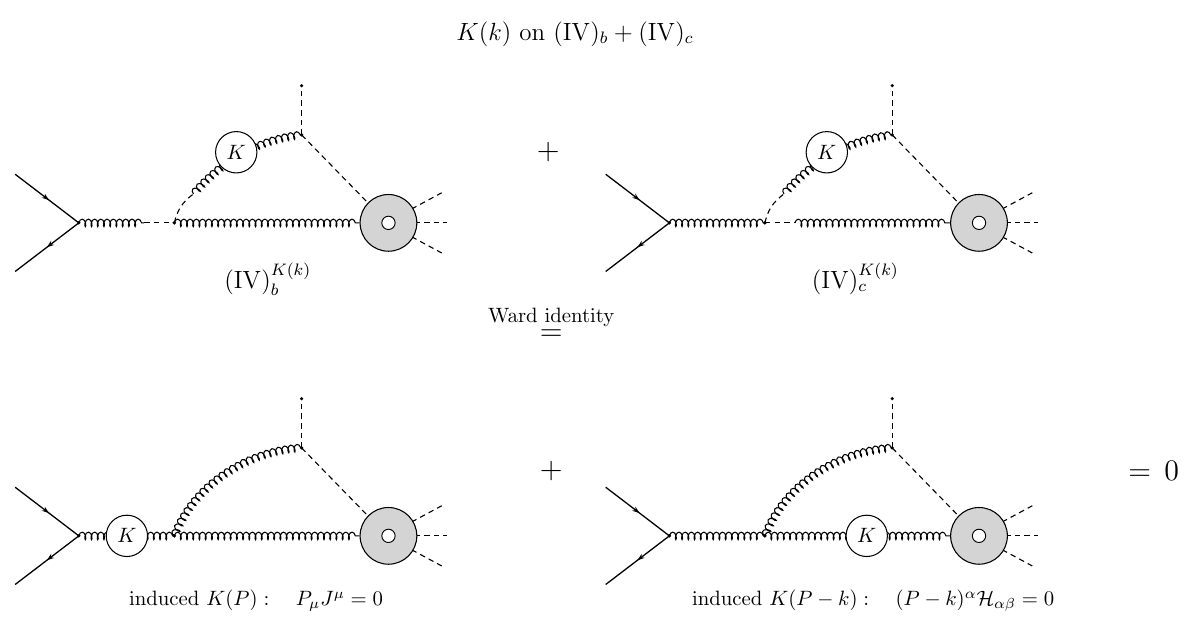}
        \caption{Ward identity of the three gluon vertex}
        \label{fig:induced-K-WI}
    \end{figure}
\end{center}

With the momenta entering the first three-gluon vertex given by $(P,-k,k-P)$, the sum of the two relevant SSG vertices is
\bea
[V_{b+c}^{abc}]_{\mu\nu\alpha}
&=&
-g_s f^{abc}
\left[
g_{\mu\nu}(P+k)_\alpha
+g_{\nu\alpha}(P-2k)_\mu
\right].
\label{eq:IV-bc-vertex}
\eea
The corresponding integrand is
\bea
\mathcal I_{\mathrm{IV},b+c}(k)
&=&
\bar v(p_2)\Gamma^a_\rho u(p_1)
D^{\rho\mu}(P)
[V_{b+c}^{abc}]_{\mu\nu\alpha}
D^{\nu\beta}(k)
\nonumber\\
&&\times
g_s f^{bxd}(2p_3-k)_\beta\,
\epsilon_{3\sigma}
D^{\alpha\gamma}(P-k)
D^{\sigma\delta}(k-p_3)
\mathcal H_{\gamma\delta}^{cd}[2,m](P-k,k-p_3).
\label{eq:IV-bc-integrand}
\eea
Let  us now check that the $K$ gluonic component of the $k$ propagator is vanishing. First, contracting $k^{\nu}$ into the vertex, we get
\bea
k^{\nu}[V^{abc}_{b+c}]_{\mu\nu\alpha}=-g_sf^{abc}[k_{\mu}(P+k)_{\alpha}+k_{\alpha}(P-2k)_{\mu}]=-g_sf^{abc}[P_{\mu}k_{\alpha}+k_{\mu}(P-k)_{\alpha}].
\eea
This identity of the three gluon vertex is shown in Fig.\ \ref{fig:induced-K-WI}
Next, we use the Ward identities,
\bea
\bar v(p_2)\Gamma^a_\mu u(p_1)P_\mu&=&0\nn \\
\mathcal H_{\gamma\delta}^{cd}[2,m](P-k,k-p_3)(P-k)_{\gamma}&=&0.
\eea
The K component of the $k$ propagator therefore vanishes in the sum of the two graphs. We may consequently replace
\bea
\mathcal I_{\mathrm{IV},b+c}^{G}(k)
&=&
\bar v(p_2)\Gamma^a_\rho u(p_1)
D^{\rho\mu}(P)
[V_{b+c}^{abc}]_{\mu\nu\alpha}
D^{G,\;\nu\beta}(k)
\nonumber\\
&&\times
g_s f^{bxd}(2p_3-k)_\beta\,
\epsilon_{3\sigma}
D^{\alpha\gamma}(P-k)
D^{\sigma\delta}(k-p_3)
\mathcal H_{\gamma\delta}^{cd}[2,m](P-k,k-p_3).
\label{eq:IV-bc-G-integrand}
\eea
without changing their sum.

\begin{figure}[htbp]
 \centering
 \includegraphics[width=\linewidth]{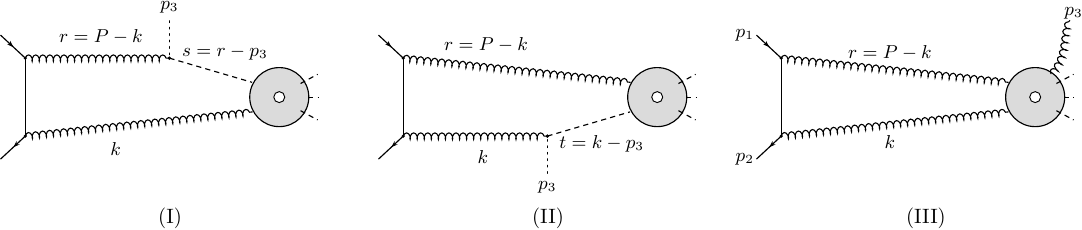}
 \caption{Graphs (I)--(III), with one momentum routing throughout.
 The external $p_3$ line is dotted at an SSG vertex and wavy
 at the ordinary gluonic attachment to the hard loop.}
 \label{fig:IR-first-three}
\end{figure}

We can now establish local finiteness . The only potentially collinear region is $k\parallel p_3$, where both $k^2$ and $(k-p_3)^2$ vanish. In this region,
\bea
(2p_3-k)_\beta G^\beta_{\ \nu}(k;q)
&=&
-2(k-p_3)_\beta G^\beta_{\ \nu}(k;q)
=O(k_\perp),
\label{eq:IV-bc-collinear}
\eea
where we used $k_\beta G^\beta_{\ \nu}(k;q)=0$. Since the two propagator denominators are $O(k_\perp^2)$, this additional transverse-momentum suppression renders the collinear integral finite.
%%%%%%%%%%%%%%%%%%%%%%%%
Continuing with our treatment of divergent graphs, we first consider the three graphs (I)--(III), with the momenta of the
upper and lower gluons entering the massless quark  as in Fig. \ref{fig:IR-first-three} line taken to be
$r=P-k$ and $k$, respectively. 

Their integrands are
\bea
\mathcal I_{\mathrm I}(k)
&=& \bar v(p_2)\Gamma^b_\rho S_0(k-p_2)\Gamma^a_\mu u(p_1)
D^{\mu\alpha}(r)D^{\rho\tau}(k)\,
\epsilon_3^\lambda
\nonumber\\
&&\times
[V_{SSG}^{acx,(23)}]_{\alpha\beta\lambda}
(r,p_3-r,-p_3)\,
D^{\beta\nu}(r-p_3)
\mathcal H^{cb}_{\nu\tau}[2,m](r-p_3,k),
\label{eq:IR-I-start}
\\
\mathcal I_{\mathrm{II}}(k)
&=& \bar v(p_2)\Gamma^b_\rho S_0(k-p_2)\Gamma^a_\mu u(p_1)
D^{\mu\nu}(r)D^{\rho\alpha}(k)\,
\epsilon_3^\lambda
\nonumber\\
&&\times
[V_{SSG}^{bcx,(23)}]_{\alpha\beta\lambda}
(k,p_3-k,-p_3)\,
D^{\beta\tau}(k-p_3)
\mathcal H^{ac}_{\nu\tau}[2,m](r,k-p_3),
\label{eq:IR-II-start}
\\
\mathcal I_{\mathrm{III}}(k)
&=& \bar v(p_2)\Gamma^b_\rho S_0(k-p_2)\Gamma^a_\mu u(p_1)
D^{\mu\nu}(r)D^{\rho\tau}(k)\,
\epsilon_3^\lambda
\mathcal H^{abx}_{\nu\tau\lambda}[3,m](r,k,-p_3).
\label{eq:IR-III-start}
\eea
The two gluons entering
$\mathcal H[3,m]$ are distinctly labelled, and the corresponding two
permutations cancel the $1/2!$ in the unlabelled graph sum.

With the quark--gluon rule $\Gamma^a_\mu=-ig_sT^a\gamma_\mu$,
the  Ward identity follows from the telescoping of the fermion loop identity
color commutator,
\bea
S_t(\ell)(-ig_sT^a\slashed r)S_t(\ell+r)
&=&g_sT^a\big[S_t(\ell)-S_t(\ell+r)\big],
\nonumber\\
g_s(T^bT^a-T^aT^b)&=&-ig_sf^{abc}T^c.
\label{eq:IR-color-WI-sign}
\eea
The corresponding non-Abelian Ward identity for the integrated hard loop is
\bea
k_1^{\mu_1}\mathcal H^{a_1a_2a_3}_{\mu_1\mu_2\mu_3}[3,m]
(k_1,k_2,k_3)
&=&-ig_sf^{a_1a_2c}\mathcal H^{ca_3}_{\mu_2\mu_3}[2,m]
(k_1+k_2,k_3)
\nonumber\\
&&-ig_sf^{a_1a_3c}\mathcal H^{a_2c}_{\mu_2\mu_3}[2,m]
(k_2,k_1+k_3).
\label{eq:IR-H3-WI-revised}
\eea
In particular, the two contractions in (III) are
\bea
r^\nu\mathcal H^{abx}_{\nu\tau\lambda}[3,m](r,k,-p_3)
&=&-ig_sf^{abc}\mathcal H^{cx}_{\tau\lambda}[2,m](P,-p_3)
-ig_sf^{axc}\mathcal H^{bc}_{\tau\lambda}[2,m](k,r-p_3),
\nonumber\\
k^\tau\mathcal H^{abx}_{\nu\tau\lambda}[3,m](r,k,-p_3)
&=&-ig_sf^{bac}\mathcal H^{cx}_{\nu\lambda}[2,m](P,-p_3)
-ig_sf^{bxc}\mathcal H^{ac}_{\nu\lambda}[2,m](r,k-p_3).
\label{eq:IR-H3-WI-rk}
\eea

\begin{figure}[ht]
\centering
\includegraphics[width=.8\linewidth]{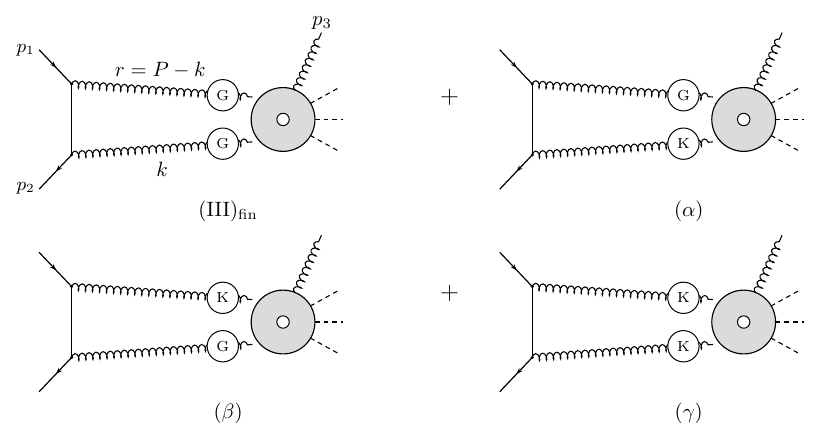}
\caption{The one-sided G--K decomposition of the two gluons entering the three-gluon hard loop of graph (III). The gluon of momentum $r=P-k$ is drawn above the gluon of momentum $k$.}
\label{fig:IR-III-four}
\end{figure}

First decompose the propagators of (III) only at their ends adjacent to
$\mathcal H[3,m]$. In the notation of Fig.~\ref{fig:IR-III-four},
\bea
\mathcal I_{\mathrm{III}}(k)
&=&\mathcal I_{\mathrm{III},\mathrm{fin}}(k)
+\mathcal I_{\mathrm{III},\alpha}(k)
+\mathcal I_{\mathrm{III},\beta}(k)
+\mathcal I_{\mathrm{III},\gamma}(k),
\nonumber\\
\mathcal I_{\mathrm{III},\mathrm{fin}}(k)
&=&\bar v(p_2)\Gamma^b_\rho S_0(k-p_2)\Gamma^a_\mu u(p_1)
D_G^{\mu\nu}(r;q)D_G^{\rho\tau}(k;q)\,
\epsilon_3^\lambda
\mathcal H^{abx}_{\nu\tau\lambda}[3,m](r,k,-p_3),
\nonumber\\
\mathcal I_{\mathrm{III},\alpha}(k)
&=&\bar v(p_2)\Gamma^b_\rho S_0(k-p_2)\Gamma^a_\mu u(p_1)
D_G^{\mu\nu}(r;q)D_K^{\rho\tau}(k;q)\,
\epsilon_3^\lambda
\mathcal H^{abx}_{\nu\tau\lambda}[3,m](r,k,-p_3),
\nonumber\\
\mathcal I_{\mathrm{III},\beta}(k)
&=&\bar v(p_2)\Gamma^b_\rho S_0(k-p_2)\Gamma^a_\mu u(p_1)
D_K^{\mu\nu}(r;q)D_G^{\rho\tau}(k;q)\,
\epsilon_3^\lambda
\mathcal H^{abx}_{\nu\tau\lambda}[3,m](r,k,-p_3),
\nonumber\\
\mathcal I_{\mathrm{III},\gamma}(k)
&=&\bar v(p_2)\Gamma^b_\rho S_0(k-p_2)\Gamma^a_\mu u(p_1)
D_K^{\mu\nu}(r;q)D_K^{\rho\tau}(k;q)\,
\epsilon_3^\lambda
\mathcal H^{abx}_{\nu\tau\lambda}[3,m](r,k,-p_3).
\label{eq:IR-III-four}
\eea
The first term is finite in the $r\parallel p_1$ and $k\parallel p_2$
regions. This follows from the collinear supression of the G gluons discussed in Sec.\ \ref{sec:transient}. The double-longitudinal contraction ($\mathcal{I}_{\gamma}$) is non zero but factorizes from the hard loop completely:
\bea
r^\nu k^\tau\mathcal H^{abx}_{\nu\tau\lambda}[3,m](r,k,-p_3)
&=&-ig_sf^{abc}\,k^\tau
\mathcal H^{cx}_{\tau\lambda}[2,m](P,-p_3).
\label{eq:IR-H3-double-WI-revised}
\eea
Finally, applying the non abelian ward identity in Eq.\ \ref{eq:IR-H3-WI-revised} to $\mathcal {I}_{\mathrm{III},\alpha/\beta}$, we get
\bea
\mathcal I_{\mathrm{III},\alpha}&=&\mathcal I_{\alpha,\text{nf}}+\mathcal I_{\alpha,\text{fact}}\nn\\
\mathcal I_{\mathrm{III},\beta}&=&\mathcal I_{\beta,\text{nf}}+\mathcal I_{\beta,\text{fact}}\nn\\
\mathcal I_{\alpha,\text{nf}}&=&-ig_sf^{bxc},
\bar v(p_2)\Gamma^b_\rho
S_0(k-p_2)\Gamma^a_\mu u(p_1)
D_G^{\mu\nu}(r;q)\nn \\&&
\frac{-i(2q+k)^\rho}
{(k^2+i0)(k^2+2q\cdot k+i0)}
\epsilon_3^\lambda
\mathcal H^{ac}{\nu\lambda}[2,m](r,k-p_3), \nn \\
\mathcal I_{\beta,\text{nf}}&=&-ig_sf^{axc},
\bar v(p_2)\Gamma^b_\rho
S_0(k-p_2)\Gamma^a_\mu u(p_1)\nonumber\\
&&
\frac{-i(2q+r)^\mu}
{(r^2+i0)(r^2+2q\cdot r+i0)}
D_G^{\rho\tau}(k;q)
\epsilon_3^\lambda
\mathcal H^{bc}_{\tau\lambda}[2,m](k,r-p_3).\label{eq:III-nf}
\eea

For graphs (I) and (II), the Abelian Ward identity of $\mathcal H[2,m]$
first permits $D(k)\to D_G(k;q)$ in (I), and $D(r)\to D_G(r;q)$ in
(II). We then decompose the remaining propagator at the SSG vertex,
\bea
\mathcal I_{\mathrm I}(k)
&=&\mathcal I_{\mathrm I,\delta}(k)+\mathcal I_{\mathrm I,a}(k),
\nonumber\\
\mathcal I_{\mathrm {II}}(k)
&=&\mathcal I_{\mathrm {II},\sigma}(k)+\mathcal I_{\mathrm {II},a}(k).
\label{eq:IR-I-II-two}
\eea
\begin{figure}[ht]
\centering
\includegraphics[width=.7\linewidth]{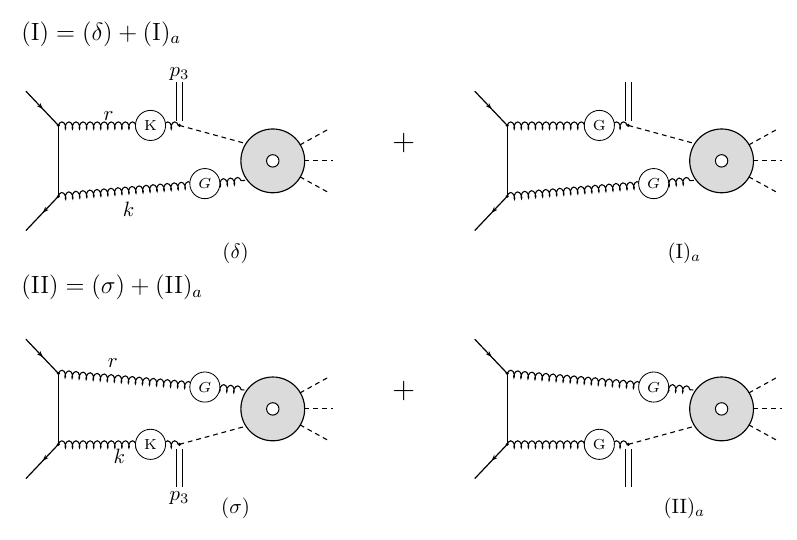}
\caption{One-sided G--K decomposition of graphs (I),(II) at the SSG ends, after replacing the other propagator by its G component using the two-gluon hard-loop Ward identity.}
\label{fig:IR-I-II-two}
\end{figure}
Here the labels $\delta$ and $\sigma$ designate the longitudinal
contributions shown in Fig.~\ref{fig:IR-I-II-two}. 
Their longitudinal contractions are
\bea
r^\alpha[V^{acx,(23)}_{SSG}]_{\alpha\beta\lambda}
(r,p_3-r,-p_3)
&=&g_sf^{acx}(r-p_3)^2 g_{\beta\lambda},
\nonumber\\
k^\alpha[V^{bcx,(23)}_{SSG}]_{\alpha\beta\lambda}
(k,p_3-k,-p_3)
&=&g_sf^{bcx}(k-p_3)^2g_{\beta\lambda}.
\label{eq:IR-SSG-WI-IK-IIK}
\eea
Inserting these equations into the two longitudinal terms gives
\bea
\mathcal I_{\mathrm I,\delta}(k)
&=&-ig_sf^{acx}\,
\bar v(p_2)\Gamma^b_\rho S_0(k-p_2)\Gamma^a_\mu u(p_1)
\frac{-i}{r^2+i0}\frac{(2q+r)^\mu}{r^2+2q\cdot r+i0}
\nonumber\\
&&\times D_G^{\rho\tau}(k;q)\,
\epsilon_3^\lambda
\mathcal H^{cb}_{\lambda\tau}[2,m](r-p_3,k),
\label{eq:IR-I-delta}
\\
\mathcal I_{\mathrm {II},\sigma}(k)
&=&-ig_sf^{bcx}\,
\bar v(p_2)\Gamma^b_\rho S_0(k-p_2)\Gamma^a_\mu u(p_1)
D_G^{\mu\nu}(r;q)
\frac{-i}{k^2+i0}\frac{(2q+k)^\rho}{k^2+2q\cdot k+i0}
\nonumber\\
&&\times\epsilon_3^\lambda
\mathcal H^{ac}_{\nu\lambda}[2,m](r,k-p_3).
\label{eq:IR-II-sigma}
\eea
The two nonfactorizing terms generated by the hard-loop Ward identity in graph III, given in Eq.\ (\ref{eq:III-nf})
cancel  these terms at fixed $k$:
\bea
\left.\mathcal I_{\mathrm{III},\beta,\text{nf}}(k)\right.+\mathcal I_{\mathrm I,\delta}(k)
&=&0,
\nonumber\\
\left.\mathcal I_{\mathrm{III},\alpha,\text{nf}}(k)\right.+\mathcal I_{\mathrm {II},\sigma}(k)
&=&0.
\label{eq:IR-III-I-II-local-cancel}
\eea

Adding all five longitudinal contributions, the entire sum reduces to
three explicitly factorized terms:
\bea
&&\mathcal I_{\mathrm{III},\alpha}(k)
+\mathcal I_{\mathrm{III},\beta}(k)
+\mathcal I_{\mathrm{III},\gamma}(k)
+\mathcal I_{\mathrm I,\delta}(k)
+\mathcal I_{\mathrm {II},\sigma}(k)
\nonumber\\
&=&ig_sf^{abc}\,
\bar v(p_2)\Gamma^b_\rho S_0(k-p_2)\Gamma^a_\mu u(p_1)
\epsilon_3^\lambda
\Bigg\{
D_G^{\mu\nu}(r;q)
\frac{-i}{k^2+i0}\frac{(2q+k)^\rho}{k^2+2q\cdot k+i0}
\mathcal H^{cx}_{\nu\lambda}[2,m](P,-p_3)
\nonumber\\
&&\hspace{1.0cm}
-\frac{-i}{r^2+i0}\frac{(2q+r)^\mu}{r^2+2q\cdot r+i0}
D_G^{\rho\tau}(k;q)
\mathcal H^{cx}_{\tau\lambda}[2,m](P,-p_3)
\nonumber\\
&&\hspace{1.0cm}
-\frac{(-i)^2}{(r^2+i0)(k^2+i0)}
\frac{(2q+r)^\mu(2q+k)^\rho}
{(r^2+2q\cdot r+i0)(k^2+2q\cdot k+i0)}
k^\tau\mathcal H^{cx}_{\tau\lambda}[2,m](P,-p_3)
\Bigg\}.
\label{eq:IR-III-I-II-mega}
\eea

These are fully factorized, exact graphical identities. 

\begin{figure}[htbp]
 \centering
 \includegraphics[width=\linewidth]{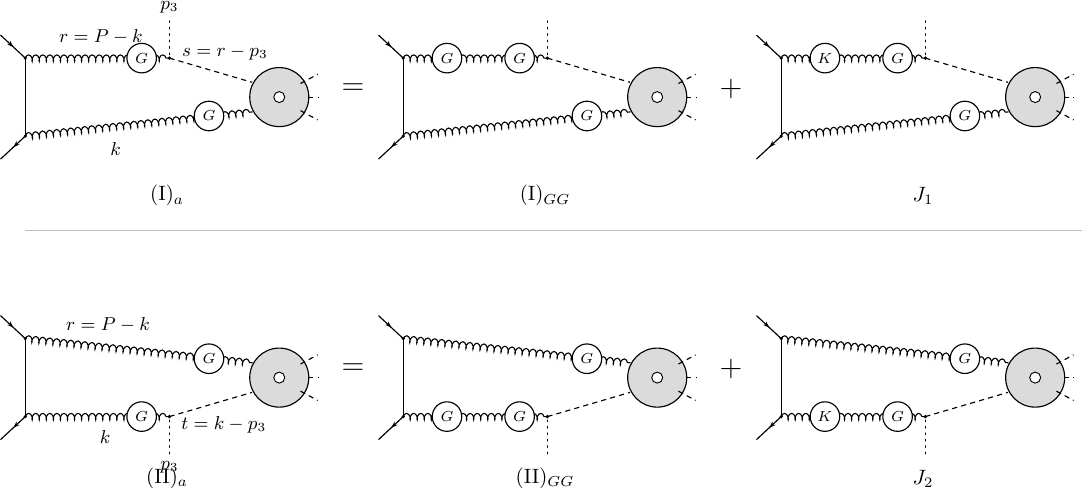}
 \caption{Second-end $G$--$K$ decomposition of $(\mathrm I)_a$
 and $(\mathrm {II})_a$.  The $G$ at the SSG vertex stays fixed;
 $J_1$ and $J_2$ have the $K$ insertion at the quark-current end.}
 \label{fig:IR-second-GK}
\end{figure}

The remaining parts of graphs (I) and (II), denoted by
$\mathcal I_{\mathrm I,a}$ and $\mathcal I_{\mathrm {II},a}$, contain a one-sided
G projector pointing toward the fermion loop as well as the triple gluon vertex for the $r$ and $k$ propagators, respectively. Explicitly they are

\bea
\mathcal I_{\mathrm I,a}(k)
&=&\bar v(p_2)\Gamma^b_\rho S_0(k-p_2)\Gamma^a_\mu u(p_1)
\frac{-i}{r^2+i0}
G^\mu_{\ \xi}(r;q)g^{\xi\alpha}
D_G^{\rho\tau}(k;q)\,\epsilon_3^\lambda
\nonumber\\
&&\times [V^{acx,(23)}_{SSG}]_{\alpha\beta\lambda}
(r,p_3-r,-p_3)D^{\beta\nu}(r-p_3)
\mathcal H^{cb}_{\nu\tau}[2,m](r-p_3,k),
\label{eq:IR-I-a}
\\
\mathcal I_{\mathrm {II},a}(k)
&=&\bar v(p_2)\Gamma^b_\rho S_0(k-p_2)\Gamma^a_\mu u(p_1)
D_G^{\mu\nu}(r;q)
\frac{-i}{k^2+i0}
G^\rho_{\ \xi}(k;q)g^{\xi\alpha}
\epsilon_3^\lambda
\nonumber\\
&&\times [V^{bcx,(23)}_{SSG}]_{\alpha\beta\lambda}
(k,p_3-k,-p_3)D^{\beta\tau}(k-p_3)
\mathcal H^{ac}_{\nu\tau}[2,m](r,k-p_3).
\label{eq:IR-II-a}
\eea
In graphs $\mathrm{I}_a,\mathrm{II}_a$,  we now decompose the gluon labeled $r,k$ respectively in their
\emph{opposite}, quark-current end.  For a gluon of momentum $u$,
the mixed term in the two sided GK is
\begin{align}
&\frac{-i}{u^2+i0}
 K_\xi{}^\mu(-u;q)g^{\xi\zeta}G_\zeta{}^\alpha(u;q)
 \nonumber\\
&\qquad=\frac{-i}{u^2+i0}
 \frac{-u^\mu(2q-u)_\zeta G^{\zeta\alpha}(u;q)}
 {u^2-2q\cdot u+i0}.
 \label{eq:IR-Kcurrent-GSSG-new}
\end{align}
In particular, the momentum at the current end is exactly $u^\mu$.
The resulting identities, shown in Fig.~\ref{fig:IR-second-GK}, are
\begin{align}
\mathcal I_{\mathrm I,a}(k)
 &=\mathcal I_{\mathrm I,GG}(k)+J_1(k),
 \nonumber\\
\mathcal I_{\mathrm {II},a}(k)
 &=\mathcal I_{\mathrm {II},GG}(k)+J_2(k).
 \label{eq:IR-J1-J2-split-new}
\end{align}

The terms labeled $\mathrm{I,II}_{GG}$ are collinear finite in all collinear limits following the arguments made in Sec.\ \ref{sec:transient}.
The two remaining longitudinal integrands are explicitly
\begin{align}
J_1(k)
={}&\bar v(p_2)\Gamma^b_\rho S_0(k-p_2)
 \Gamma^a_\mu u(p_1)
 D_G^{\rho\tau}(k;q)
 \nonumber\\[-2pt]
&\times\frac{-i}{r^2+i0}
 K_\xi{}^\mu(-r;q)g^{\xi\zeta}G_\zeta{}^\alpha(r;q)
 \nonumber\\[-2pt]
&\times\epsilon_3^\lambda
 [V_{SSG}^{acx,(23)}]_{\alpha\beta\lambda}(r,-s,-p_3)
 D^{\beta\nu}(s)\,
 \mathcal H^{cb}_{\nu\tau}[2,m](s,k),
 \label{eq:IR-J1-new}\\[5pt]
J_2(k)
={}&\bar v(p_2)\Gamma^b_\rho S_0(k-p_2)
 \Gamma^a_\mu u(p_1)
 D_G^{\mu\nu}(r;q)
 \nonumber\\[-2pt]
&\times\frac{-i}{k^2+i0}
 K_\xi{}^\rho(-k;q)g^{\xi\zeta}G_\zeta{}^\alpha(k;q)
 \nonumber\\[-2pt]
&\times\epsilon_3^\lambda
 [V_{SSG}^{bcx,(23)}]_{\alpha\beta\lambda}(k,-t,-p_3)
 D^{\beta\tau}(t)\,
 \mathcal H^{ac}_{\nu\tau}[2,m](r,t).
 \label{eq:IR-J2-new}
\end{align}
The explicit expressions for $\mathrm{I,II}_{GG}$ are obtained by replacing $K_\xi{}^\mu(-r;q)$, $K_\xi{}^\rho(-k;q)$ by $G_\xi{}^\mu(-r;q)$, $G_\xi{}^\rho(-k;q)$ in Eqs.\ (\ref{eq:IR-J1-new}),(\ref{eq:IR-J2-new}) respectively. 
Having made this decomposition, we will find that the graphs $J_1,J_2$ will cancel a third term coming from graph $\mathrm{IV}_a$ which we now turn to. 

\begin{figure}[htbp]
 \centering
 \includegraphics[width=\linewidth]{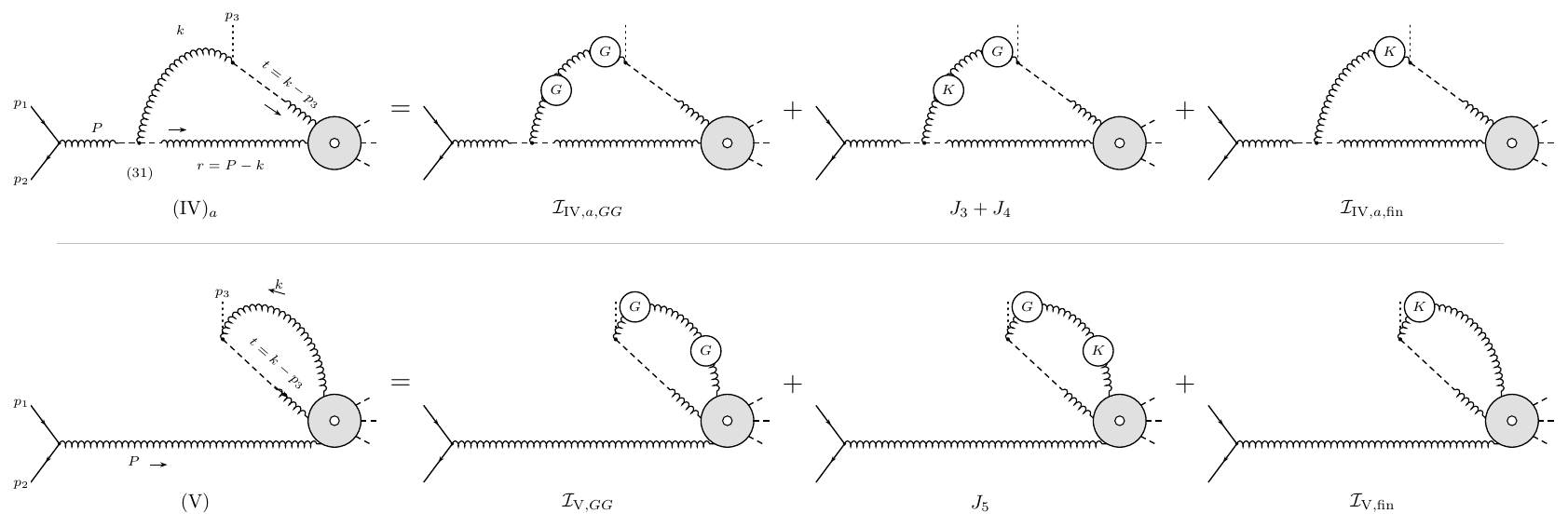}
 \caption{The exact three-sector decomposition of $(\mathrm{IV})_a$
 and (V).  The $K-G$ terms produce
 $J_3,J_4$ and $J_5$.  The two other sectors in each row
 are locally infrared finite.}
 \label{fig:IR-IV-V-splits}
\end{figure}

For these graphs we use precisely the same routing $r=P-k$,
$t=k-p_3$.  The relevant vertices in $(\mathrm{IV})_a$ are
\begin{align}
[V^{abc,(31)}]_{\mu\nu\alpha}(P,-k,-r)
 &=g_sf^{abc}g_{\mu\alpha}(2P-k)_\nu,
 \nonumber\\
[V^{bdx,(23)}]_{\beta\gamma\lambda}(k,-t,-p_3)
 &=-g_sf^{bdx}g_{\gamma\lambda}(2p_3-k)_\beta.
 \label{eq:IR-IV-V-vertices-new}
\end{align}
The complete integrands are
\begin{align}
\mathcal I_{\mathrm{IV},a}(k)
={}&\bar v(p_2)\Gamma^a_\rho u(p_1)
 D^{\rho\mu}(P)
 [V^{abc,(31)}]_{\mu\nu\alpha}(P,-k,-r)
 \nonumber\\[-2pt]
&\times D^{\nu\beta}(k)D^{\alpha\alpha'}(r)
 \epsilon_3^\lambda
 [V^{bdx,(23)}]_{\beta\gamma\lambda}(k,-t,-p_3)
 \nonumber\\[-2pt]
&\times D^{\gamma\gamma'}(t)
 \mathcal H^{cd}_{\alpha'\gamma'}[2,m](r,t),
 \label{eq:IR-IVa-integrand-new}\\[5pt]
\mathcal I_{\mathrm V}(k)
={}&\bar v(p_2)\Gamma^a_\rho u(p_1)
 D^{\rho\mu}(P)
 D^{\beta\beta'}(k)D^{\gamma\gamma'}(t)
 \nonumber\\[-2pt]
&\times\epsilon_3^\lambda
 [V^{bdx,(23)}]_{\beta\gamma\lambda}(k,-t,-p_3)
 \mathcal H^{abd}_{\mu\beta'\gamma'}[3,m](P,-k,t).
 \label{eq:IR-V-integrand-new}
\end{align}
These topologies and their projector assignments are shown in
Fig.~\ref{fig:IR-IV-V-splits}.

The $k$ line is now decomposed first at its scalar-scalar-gluon end and then,
only for its $G$ component, at the other end in both graphs:
\begin{align}
D^{\nu\beta}(k)
={}&\frac{-i}{k^2+i0}
 \Big[
   K_\xi{}^\nu(-k;q)g^{\xi\zeta}G_\zeta{}^\beta(k;q)
 \nonumber\\[-2pt]
&\qquad
  +G_\xi{}^\nu(-k;q)g^{\xi\zeta}G_\zeta{}^{\beta}(k;q)
  +g^{\nu\zeta}K_\zeta{}^\beta(k;q)
 \Big].\nn \\
 \label{eq:IR-IVV-projectors-new}
D^{\beta\beta'}(k)
={}&\frac{-i}{k^2+i0}
 \Big[
   K_\xi{}^{\beta'}(-k;q)g^{\xi\zeta}G_\zeta{}^{\beta}(k;q)
 \nonumber\\[-2pt]
&\qquad
  +G_\xi{}^{\beta'}(-k;q)g^{\xi\zeta}G_\zeta{}^{\beta}(k;q)
  +g^{\beta'\zeta}K_\zeta{}^{\beta}(k;q)
 \Big].
\end{align}

Let us first write down explicit expressions for the locally finite $GG,\text{fin}$ terms in Figs.\ \ref{fig:IR-IV-V-splits}.
\bea
\mathcal I_{\mathrm{IV},a,GG}(k)
&=&\bar v(p_2)\Gamma^a_\rho u(p_1)
 D^{\rho\mu}(P)
 [V^{abc,(31)}]_{\mu\nu\alpha}(P,-k,-r)D^{\gamma\gamma'}(t)
 \mathcal H^{cd}_{\alpha'\gamma'}[2,m](r,t),
 \nonumber\\[-2pt]
&&\times G_\xi{}^\nu(-k;q)g^{\xi\zeta}G_\zeta{}^{\beta}(k;q)D^{\alpha\alpha'}(r)
 \epsilon_3^\lambda
 [V^{bdx,(23)}]_{\beta\gamma\lambda}(k,-t,-p_3)
 \label{eq:IR-IVa-integrand_GG}\\
 \mathcal I_{\mathrm{IV},a,fin}(k)
&=&\bar v(p_2)\Gamma^a_\rho u(p_1)
 D^{\rho\mu}(P)
 [V^{abc,(31)}]_{\mu\nu\alpha}(P,-k,-r)D^{\gamma\gamma'}(t)
 \mathcal H^{cd}_{\alpha'\gamma'}[2,m](r,t),
 \nonumber\\[-2pt]
&&\times g^{\nu\zeta}K_\zeta{}^\beta(k;q)D^{\alpha\alpha'}(r)
 \epsilon_3^\lambda
 [V^{bdx,(23)}]_{\beta\gamma\lambda}(k,-t,-p_3).
 \label{eq:IR-IVa-integrand_fin}
\eea
The first GG piece is IR finite following the standard G gluon collinear suppression encountered in Sec.\ \ref{sec:transient}.  The second expression is IR finite because the contraction $k^{\beta}[V^{dbx}]_{\beta\gamma\lambda}(k,-t-p_3)\propto t^2 $ which cancels the $t^2$ denominator. As a result, in the $t\parallel p_3$ limit, the previously logarithmic divergence is now power counting convergent. 

Similarly for $V$, we have
\bea
\mathcal I_{\mathrm V,GG}(k)
&=&\bar v(p_2)\Gamma^a_\rho u(p_1)
 D^{\rho\mu}(P)
 D^{\gamma\gamma'}(t)\mathcal H^{abd}_{\mu\beta'\gamma'}[3,m](P,-k,t)
 \nonumber\\[-2pt]
&&\times\epsilon_3^\lambda
 [V^{bdx,(23)}]_{\beta\gamma\lambda}(k,-t,-p_3)G_\xi{}^{\beta'}(-k;q)g^{\xi\zeta}G_\zeta{}^{\beta}(k;q),\\
 \mathcal I_{\mathrm V,fin}(k)
&=&\bar v(p_2)\Gamma^a_\rho u(p_1)
 D^{\rho\mu}(P)
 D^{\gamma\gamma'}(t)\mathcal H^{abd}_{\mu\beta'\gamma'}[3,m](P,-k,t)
 \nonumber\\[-2pt]
&&\times\epsilon_3^\lambda
 [V^{bdx,(23)}]_{\beta\gamma\lambda}(k,-t,-p_3)g^{\beta'\zeta}K_\zeta{}^{\beta}(k;q),
\eea
In much the same way as $\mathrm{IV}_a$, these pieces are also locally finite because in $\mathcal I_{\mathrm V,GG}(k)$ the G tensors are power suppressed near the logarithmically divergent pinch $k\parallel p_3$ and $ \mathcal I_{\mathrm V,fin}(k)$ is finite because $k^{\beta}[V^{bdx}]_{\beta\gamma\lambda}(k,-t-p_3)\propto t^2 $ as before.

Turning to what remains, in the $J_3+J_4$ terms in Fig. \ref{fig:IR-IV-V-splits}, the $KG$ term admits a vertex factor 
\begin{equation}
 k^\nu[V^{abc,(31)}]_{\mu\nu\alpha}(P,-k,-r)
 =g_sf^{abc}(P^2-r^2)g_{\mu\alpha}.
 \label{eq:IR-IV-firstWI-new}
\end{equation}
Since
\begin{equation}
 (P^2-r^2)D^{\rho\mu}(P)g_{\mu\alpha}
 D^{\alpha\alpha'}(r)
 =-iD^{\rho\alpha'}(r)+iD^{\rho\alpha'}(P),
 \label{eq:IR-IV-collapse-new}
\end{equation}
we obtain $J_3$ and $J_4$ separately:
\begin{align}
J_3(k)
={}&-ig_sf^{abc}\,
 \bar v(p_2)\Gamma^a_\rho u(p_1)D^{\rho\alpha'}(r)
 \nonumber\\[-2pt]
&\times\frac{-i}{k^2+i0}
 \frac{-(2q-k)_\zeta G^{\zeta\beta}(k;q)}
      {k^2-2q\cdot k+i0}
 \nonumber\\[-2pt]
&\times\epsilon_3^\lambda
 [V_{SSG}^{bdx,(23)}]_{\beta\gamma\lambda}(k,-t,-p_3)
 D^{\gamma\gamma'}(t)
 \mathcal H^{cd}_{\alpha'\gamma'}[2,m](r,t),
 \label{eq:IR-J3-new}\\[5pt]
J_4(k)
={}&+ig_sf^{abc}\,
 \bar v(p_2)\Gamma^a_\rho u(p_1)D^{\rho\alpha'}(P)
 \nonumber\\[-2pt]
&\times\frac{-i}{k^2+i0}
 \frac{-(2q-k)_\zeta G^{\zeta\beta}(k;q)}
      {k^2-2q\cdot k+i0}
 \nonumber\\[-2pt]
&\times\epsilon_3^\lambda
 [V_{SSG}^{bdx,(23)}]_{\beta\gamma\lambda}(k,-t,-p_3)
 D^{\gamma\gamma'}(t)
 \mathcal H^{cd}_{\alpha'\gamma'}[2,m](r,t).
 \label{eq:IR-J4-new}
\end{align}

\begin{figure}[htbp]
 \centering
 \includegraphics[width=\linewidth]{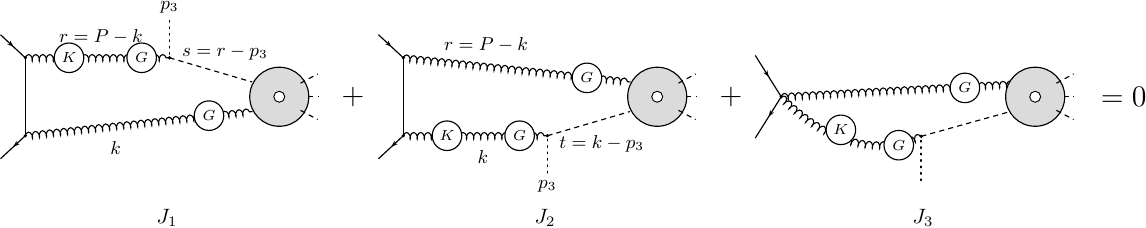}
 \caption{The Ward identity at the massless fermion current.}
 \label{fig:IR-J123-cancel}
\end{figure}

Let us now argue that $J_1+J_2+J_3=0$, as an exact ward identity. First, we utilize the abelian Ward identity to replace in Eq.\ (\ref{eq:IR-J3-new}), $D^{\rho\alpha'}(r)$ by $D_G^{\rho\alpha'}(r)$.  
In $J_1$ we shift $k\mapsto P-k$, relabel the dummy color
indices, and use Bose symmetry of the integrated two-gluon
hard loop.  Both quark-line attachments then carry the same
longitudinal momentum $k$, and
\begin{align}
&k^\rho\bar v(p_2)
 \Big[\Gamma^a_\mu S_0(p_1-k)\Gamma^b_\rho
       +\Gamma^b_\rho S_0(k-p_2)\Gamma^a_\mu\Big]u(p_1)
 \nonumber\\
&\qquad=g_s\bar v(p_2)[T^b,\Gamma^a_\mu]u(p_1)
 =ig_sf^{bac}\bar v(p_2)\Gamma^c_\mu u(p_1).
 \label{eq:IR-quark-color-WI-new}
\end{align}
 Utilizing Eq.\ (\ref{eq:IR-quark-color-WI-new}) in Eqs.\ (\ref{eq:IR-J1-new}),(\ref{eq:IR-J2-new}) and combing with Eq.\ (\ref{eq:IR-J3-new}), we find 
\begin{equation}
 {J_1(P-k)+J_2(k)+J_3(k)=0.}
 \label{eq:IR-J123-zero-new}
\end{equation}
This ward identity is shown in Fig.\ \ref{fig:IR-J123-cancel}. 

The remaining $KG$ contribution to (V) is
\begin{align}
J_5(k)
={}&\bar v(p_2)\Gamma^a_\rho u(p_1)D^{\rho\mu}(P)
 \nonumber\\[-2pt]
&\times\frac{-i}{k^2+i0}
 \frac{-(2q-k)_\zeta G^{\zeta\beta}(k;q)}
      {k^2-2q\cdot k+i0}
 \nonumber\\[-2pt]
&\times\epsilon_3^\lambda
 [V_{SSG}^{bdx,(23)}]_{\beta\gamma\lambda}(k,-t,-p_3)
 D^{\gamma\gamma'}(t)
 \nonumber\\[-2pt]
&\times k^{\beta'}
 \mathcal H^{abd}_{\mu\beta'\gamma'}[3,m](P,-k,t).
 \label{eq:IR-J5-new}
\end{align}
The three-gluon hard-loop Ward identity gives
\begin{align}
&k^{\beta'}\mathcal H^{abd}_{\mu\beta'\gamma'}[3,m]
 (P,-k,t)
 \nonumber\\[-2pt]
&\quad=-ig_sf^{abc}
 \mathcal H^{cd}_{\mu\gamma'}[2,m](r,t)
      +ig_sf^{bdc}
 \mathcal H^{ac}_{\mu\gamma'}[2,m](P,-p_3).
 \label{eq:IR-V-WI-new}
\end{align}
Its first term equals $-J_4(k)$, with exactly the same
SSG projector factor, whereas the second has a hard loop
independent of $k$.  Thus the nonfactorizing pieces cancel:
\begin{align}
J_4(k)+J_5(k)
={}&ig_sf^{bdc}\,
 \bar v(p_2)\Gamma^a_\rho u(p_1)D^{\rho\mu}(P)
 \nonumber\\[-2pt]
&\times\frac{-i}{k^2+i0}
 \frac{-(2q-k)_\zeta G^{\zeta\beta}(k;q)}
      {k^2-2q\cdot k+i0}
 \nonumber\\[-2pt]
&\times\epsilon_3^\lambda
 [V_{SSG}^{bdx,(23)}]_{\beta\gamma\lambda}(k,-t,-p_3)
 D^{\gamma\gamma'}(t)
 \mathcal H^{ac}_{\mu\gamma'}[2,m](P,-p_3).
 \label{eq:IR-J45-factor-raw-new}
\end{align}
Using $f^{bdc}f^{bdx}=C_A\delta^{cx}$, the explicitly
factorized result is
\begin{align}
J_4(k)+J_5(k)
={}&-ig_s^2C_A
 \frac{(2q-k)_\zeta G^{\zeta\beta}(k;q)(2p_3-k)_\beta}
 {(k^2+i0)(t^2+i0)(k^2-2q\cdot k+i0)}
 \\[2pt]
&\times\bar v(p_2)\Gamma^a_\rho u(p_1)D^{\rho\mu}(P)
 \epsilon_3^{\gamma'}
 \mathcal H^{ax}_{\mu\gamma'}[2,m](P,-p_3).
 \label{eq:IR-J45-factor-final-new}
\end{align}
The hard tensor on the right has no loop-momentum dependence.
This gives a genuine factorized correction to the external $p_3$ line.

The $GG$ contributions of (I) and (II) are collinear finite,
but retain their respective genuine soft singularities at
$r\to0$ and $k\to0$.  These are removed by the local soft
counterterms which we discuss in conclusion to this section.  All longitudinal contributions
from (I)--(V) have meanwhile either canceled or reduced to
the factorized expressions in Eqs.~(\ref{eq:IR-III-I-II-mega})
and (\ref{eq:IR-J45-factor-final-new}).

Let us finally turn to the soft counterterms to the graphs $\mathrm{I}_{GG},\mathrm{II}_{GG}$. To construct these, we simply substitute the relevant soft limit in the integrand. 

In the limit $r\to0$, the massless quark propagator and the
SSG vertex in graph (I) satisfy
\begin{align}
S_0(p_1-r)\Gamma^a_\mu u(p_1)
&=
-g_sT^a
\frac{-2p_{1\mu}}{(p_1- r)^2+i0}\,
u(p_1)+O(r^0),
\nonumber\\
\epsilon_3^\lambda
[V_{SSG}^{acx,(23)}]_{\alpha\beta\lambda}
(r,-s,-p_3)D^{\beta\nu}(s)
&=
-ig_sf^{acx}
\frac{-2p_{3\alpha}}{(p_3- r)^2+i0}\,
\epsilon_3^\nu+O(r^0).
\label{eq:IR-I-soft-approx}
\end{align}
Similarly, in the limit $k\to0$, graph (II) satisfies
\begin{align}
\bar v(p_2)\Gamma^b_\rho S_0(k-p_2)
&=
g_s\frac{-p_{2\rho}}{(p_2- k)^2+i0}\,
\bar v(p_2)T^b+O(k^0),
\nonumber\\
\epsilon_3^\lambda
[V_{SSG}^{bcx,(23)}]_{\alpha\beta\lambda}
(k,-t,-p_3)D^{\beta\tau}(t)
&=
-ig_sf^{bcx}
\frac{-2p_{3\alpha}}{(p_3- k)^2+i0}\,
\epsilon_3^\tau+O(k^0).
\label{eq:IR-II-soft-approx}
\end{align}

In either limit, the massive fermion loop reduces to
$\mathcal H[2,m](P,-p_3)$, and the remaining hard gluon
propagator can be replaced by its unprojected counterpart
using the Abelian Ward identity.

Writing
$\mathcal H^{ab}_{\mu\nu}[2,m]
=\delta^{ab}\mathcal H_{\mu\nu}[2,m]$,
the corresponding one-loop amplitude is
\begin{equation}
\mathcal M^{(1)}
=
\bar v(p_2)\Gamma^x_\rho u(p_1)\,
D^{\rho\nu}(P)\,
\epsilon_3^\lambda
\mathcal H_{\nu\lambda}[2,m](P,-p_3).
\label{eq:IR-hard-one-loop-soft}
\end{equation}

Using the color identities
\begin{align}
f^{abx}T^bT^a
&=-\frac{iC_A}{2}T^x,
&
f^{bax}T^bT^a
&=\frac{iC_A}{2}T^x,
\end{align}
we obtain the soft counterterms
\begin{align}
\mathcal C_{\mathrm I}^{\mathrm{soft}}(k)
={}&
\frac{g_s^2 C_A}{2}\,
\frac{-i}{r^2+i0}
\nonumber\\[-2pt]
&\times
\frac{
  (2p_1-r)_\mu
  G_\xi{}^\mu(-r;q)\,
  g^{\xi\zeta}\,
  G_\zeta{}^\alpha(r;q)
  (2p_3-r)_\alpha
}{
  \bigl((p_1-r)^2+i0\bigr)
  \bigl((p_3-r)^2+i0\bigr)
}
\,\mathcal M^{(1)},
\label{eq:IR-I-soft-CT}
\\[8pt]
\mathcal C_{\mathrm{II}}^{\mathrm{soft}}(k)
={}&
\frac{g_s^2 C_A}{2}\,
\frac{-i}{k^2+i0}
\nonumber\\[-2pt]
&\times
\frac{
  (2p_2-k)_\rho
  G_\xi{}^\rho(-k;q)\,
  g^{\xi\zeta}\,
  G_\zeta{}^\alpha(k;q)
  (2p_3-k)_\alpha
}{
  \bigl((p_2-k)^2+i0\bigr)
  \bigl((p_3-k)^2+i0\bigr)
}
\,\mathcal M^{(1)}.
\label{eq:IR-II-soft-CT}
\end{align}

We define the subtracted integrands by
\begin{align}
\mathcal I_{\mathrm I,GG}^{\mathrm{sub}}(k)
&=
\mathcal I_{\mathrm I,GG}(k)
-\mathcal C_{\mathrm I}^{\mathrm{soft}}(k),
\nonumber\\
\mathcal I_{\mathrm{II},GG}^{\mathrm{sub}}(k)
&=
\mathcal I_{\mathrm{II},GG}(k)
-\mathcal C_{\mathrm{II}}^{\mathrm{soft}}(k).
\label{eq:IR-GG-soft-subtractions}
\end{align}
By construction, these counterterms reproduce the
leading soft behavior at $r=0$ and $k=0$, respectively.
They remove the remaining soft logarithms without
altering the previously established collinear finiteness
of the $GG$ contributions.
In this section we choose \(q=P=p_1+p_2\). Since \(P\) is timelike,
the auxiliary denominator \(k^2+2P\cdot k\) is nonzero on every
nonzero collinear momentum \(k=xp_i\). For the momentum routings in
Eqs.~\eqref{eq:IR-III-I-II-mega}--\eqref{eq:IR-J45-factor-final-new},
the only additional pinch allowed by the Landau equations occurs when
the two denominators \(k^2\pm2P\cdot k+i0\) on the same gluon line
trap the contour at \(k=0\). In \(\mathrm{IV}_{a,GG}\), the vertex
factor \(2P-k\) is annihilated by \(G(-k;P)\), so this term vanishes.
In \(\mathrm{I}_{GG}\) and \(\mathrm{II}_{GG}\), contraction with the
longitudinal momentum cancels the adjacent propagators, and the soft
counterterms reproduce the remaining numerator at zero momentum since
\(J\cdot P=0\). In \(\mathrm{V}_{GG}\), the same contraction cancels
the \((k-p_3)^2\) propagator and leaves an explicit \(k\) contracted
with the hard tensor. Averaging each finite remainder under four-momentum
reversal (\(r\to-r\) in \(\mathrm{I}\), \(k\to-k\) otherwise) therefore
leaves a numerator of order \(|\mathbf k|^2\) at the auxiliary pinch.
In the \(P\) rest frame the scalar residue is of order
\(|\mathbf k|^{-4}\), so the remaining three-momentum integral is
locally convergent.

In this section, we have studied the graphs $\mathrm{I}-\mathrm{V}$. In each graph, we have chosen some propagators to decompose on one or both ends using the identity $G+K=1$. In each case, we found that the graphs obtained after Ward identity cancellations were either locally finite, factorized into products of one loop graphs or divergent in the soft limit and handled using a soft counterterm. Combined with the results of Sec.\ \ref{sec:transient} every genuine two loop graph that contributes to the amplitude has been rendered locally finite.

\section{Conclusions}
\label{sec:conclusions}
In this paper, we have studied local infrared finiteness for the process $q+\bar{q}\rightarrow g+nH$ (and by extension its crossed processes). We have classified graphs into topologically factorized, IR finite, transient singular and IR divergent.  Some finite graphs were then added to the set of IR divergent graphs to manifest Ward identities. 

Here, we have used two tools the scalar-scalar-gluon decomposition as well as the one-sided and  two-sided G-K decomposition. For transient singularities, we showed that it is possible to make every graph locally infrared finite by a non-local counter-term (which only integrates to zero after summing over its own gauge invariant subset). This was constructed utilizing the one-sided G-K decomposition and the  Ward identity that is only satisfied after the permutation sum on the fermion loop and its loop integral is carried out.  The key observation here was that it is possible to replace a gluon by its G counterpart when only two gluons attach to the fermion loop. We did not need shift counterterms that are often used in other approaches to local finiteness. 

For IR divergent graphs, we observed that Feynman graphs after the two-sided GK decomposition, satisfy exact Ward identities which factorize longitudinal K gluons into products of one loop graphs. The physical G gluonic propagation is always collinear finite due to the identity in Eq.\ (\ref{eq:G-Kpow}) while soft divergence was identified with gauge invariance and gauge freedom. We constructed explicit soft counter terms which are easy to read off for the processes studied here. 

Let us explain the new features of this decomposition. First, we have studied colored particles in both the initial and final state. We have supplied a modification of the Feynman rules for graphs that have genuinely two-loop topologies, when replacing a gluon with its physical polarizations. Unless the polarizations are chosen to be lightlike, such a procedure is bound to leave behind soft singularities. We constructed soft counter-terms for such topologies while we retained time-like reference vectors. This opens the door to include more colored particles, particularly to include incoming and outgoing gluons as well as to systematically generate counter-terms in an automated fashion. It would be interesting to use these tools to attack the three loop form factor which resisted local factorization in \cite{Haindl:2025jte}. 
\acknowledgments
I have utilized OpenAI's ChatGPT 6 to draw the figures used in this paper. I have used it to organize the one loop integrals needed to carry out the checks described here. I have checked in each case that the integral has been correctly transcribed. I have also benefited from Feynrules and Feynarts as described in \cite{Hahn:2000kx} to generate graphs to verify the completeness of the graph list as well as to inspect the symmetry factors and integrands of various graphs.
\appendix{
\section{Checks}
\label{sec:check}
We evaluate the four contributions to the infrared poles of
$\mathcal M^{(2)}$. We use $d=4-2\epsilon$ and the
$\overline{\mathrm{MS}}$ loop measure. The scalar integrals we need are
\begin{align}
 B(a,b)&=(\mu^2e^{\gamma_E})^\epsilon\!
 \int\frac{d^d k}{i\pi^{d/2}}
 \frac{1}{(k^2-2a\cdot k+i0)(k^2-2b\cdot k+i0)},\nonumber\\
 T(a,b)&=(\mu^2e^{\gamma_E})^\epsilon\!
 \int\frac{d^d k}{i\pi^{d/2}}
 \frac{1}{(k^2+i0)(k^2-2a\cdot k+i0)(k^2-2b\cdot k+i0)} .
\end{align}
For lightlike $p,p'$ and future timelike $q$, their triangle poles are
\begin{align}
 \left.4p\cdot p'\,T(p,p')\right|_{\rm poles}
 &=-\frac{2}{\epsilon^2}
   +\frac{2}{\epsilon}\ln\frac{2p\cdot p'-i0}{\mu^2},\nonumber\\
 \left.4p\cdot(\pm q)\,T(p,\pm q)\right|_{\rm poles}
 &=-\frac{1}{\epsilon^2}
   +\frac{2}{\epsilon}\ln\frac{\pm2p\cdot q-i0}{\mu\sqrt{q^2}},\nonumber\\
 \left.4q^2T(q,-q)\right|_{\rm poles}&=-\frac{2}{\epsilon}.
 \label{eq:check-triangles}
\end{align}
The non-scaleless bubbles below have pole $1/\epsilon$.
The on-shell bubble $B(0,p)$ vanishes in dimensional regularization.
Its UV and IR residues are $+1$ and $-1$, respectively.

\textit{Five longitudinal terms.}
Consider the three terms on the right of Eq.~\eqref{eq:IR-III-I-II-mega}
in the limit $k\parallel p_2$. At leading power we set $k=xp_2$ in
the fermion numerator and in the hard propagator carrying momentum
$r=P-k$. Since $r^2=(1-x)P^2$, the Dirac equation gives
\begin{align}
 \left.
 \frac{\bar v(p_2)\gamma_\rho(\slashed k-\slashed p_2)
              \gamma_\mu u(p_1)}{r^2}
 \right|_{k=xp_2}
 &=
 \frac{-2(1-x)p_{2\rho}\,\bar v(p_2)\gamma_\mu u(p_1)}
      {(1-x)P^2}
 =
 -2p_{2\rho}\,\frac{\bar v(p_2)\gamma_\mu u(p_1)}{P^2}.
\end{align}
The second term vanishes at this order because
$p_{2\rho}G^{\rho\tau}(k;q)=0$ at $k=xp_2$. In the first term,
expand $G^{\mu\nu}(r;q)$ on the hard line. Its longitudinal part
cancels the third term by
$r^\nu\mathcal H_{\nu\lambda}[2,m](P,-p_3)
=-k^\nu\mathcal H_{\nu\lambda}[2,m](P,-p_3)$.
The metric part of $G^{\mu\nu}(r;q)$ then joins the quark current
to the hard loop with the Born propagator $1/P^2$, giving
$\mathcal M^{(1)}$ of Eq.~\eqref{eq:IR-hard-one-loop-soft}.

The remaining numerator is $-2p_2\cdot(2q+xp_2)=-4p_2\cdot q$.
Keeping the full $k$ dependence of the three remaining denominators,
we obtain the scalar triangle
\begin{align}
 -4p_2\cdot q\,T(p_2,-q)
 &=(\mu^2e^{\gamma_E})^\epsilon
 \int\frac{d^d k}{i\pi^{d/2}}\,
 \frac{-4p_2\cdot q}
 {(k^2+i0)((k-p_2)^2+i0)(k^2+2q\cdot k+i0)} .
\end{align}
This triangle also reproduces the leading soft integrand at $k=0$,
where $r=P$. The region $r\parallel p_1$ gives the corresponding
triangle $-4p_1\cdot q\,T(p_1,-q)$. Restoring the colour and coupling
factors, we find
\begin{align}
 &\left[\int\frac{d^d k}{(2\pi)^d}
 \big(\mathcal I_{\mathrm{III},\alpha}
     +\mathcal I_{\mathrm{III},\beta}
     +\mathcal I_{\mathrm{III},\gamma}
     +\mathcal I_{\mathrm I,\delta}
     +\mathcal I_{\mathrm{II},\sigma}\big)\right]_{\rm poles}
 \nonumber\\
 &\qquad =
 \left[\frac{g_s^2 C_A}{32\pi^2}\,\mathcal M^{(1)}
 \sum_{i=1}^2(-4p_i\cdot q)\,T(p_i,-q)\right]_{\rm poles}.
 \label{eq:check-five}
\end{align}

Consider next $J_4+J_5$ in Eq.~\eqref{eq:IR-J45-factor-final-new}.
The hard gluon carries momentum $P$, and the two momenta entering the
fermion loop are $P$ and $-p_3$. The quark current, the hard propagator
and the fermion loop therefore give precisely $\mathcal M^{(1)}$ of
Eq.~\eqref{eq:IR-hard-one-loop-soft}. This factorization is exact:
\begin{align}
 J_4(k)+J_5(k)
 &=-ig_s^2 C_A\,\mathcal M^{(1)}
 \frac{(2q-k)_\zeta G^{\zeta\beta}(k;q)(2p_3-k)_\beta}
 {(k^2+i0)((k-p_3)^2+i0)(k^2-2q\cdot k+i0)} .
\end{align}
It remains to integrate the scalar factor. Expanding $G(k;q)$ and
using $k\cdot(2p_3-k)=-(k-p_3)^2$, we obtain
\begin{align}
 (2q-k)_\zeta G^{\zeta\beta}(k;q)(2p_3-k)_\beta
 &=4q\cdot p_3+(k^2-2q\cdot k)+(k-p_3)^2-k^2
 \nonumber\\
 &\quad+\frac{(4q^2-k^2)(k-p_3)^2}{k^2+2q\cdot k+i0}.
\end{align}
Cancelling the corresponding propagators gives
\begin{align}
 \int\frac{d^d k}{(2\pi)^d}\,[J_4(k)+J_5(k)]
 &=\frac{g_s^2 C_A}{16\pi^2}\,\mathcal M^{(1)}
 \big[4q\cdot p_3\,T(p_3,q)+B(0,p_3)
 \nonumber\\
 &\qquad+B(0,q)-B(p_3,q)
       +4q^2T(q,-q)-B(q,-q)\big].
\end{align}
The on-shell bubble $B(0,p_3)$ vanishes, and the poles of $B(0,q)$
and $B(p_3,q)$ cancel. The last two terms give $-2/\epsilon$ and
$-1/\epsilon$, respectively. Inserting the pole of $T(p_3,q)$ from
Eq.~\eqref{eq:check-triangles}, we find
\begin{align}
 \left[\int\frac{d^d k}{(2\pi)^d}\,[J_4(k)+J_5(k)]\right]_{\rm poles}
 &=
 \left[\frac{g_s^2 C_A}{16\pi^2}\,\mathcal M^{(1)}
 \left(
 -\frac{1}{\epsilon^2}
 +\frac{2}{\epsilon}\ln\frac{2q\cdot p_3}{\mu\sqrt{q^2}}
 -\frac{3}{\epsilon}
 \right)\right]_{\rm poles}.
 \label{eq:check-J45}
\end{align}

The three topologically factorized vertex graphs give the one-loop
$q\bar q g^*$ vertex with $p_1^2=p_2^2=0$ and gluon virtuality $P^2$.
Using the massless vertex in Feynman gauge
\cite{Davydychev:2000rt}
and including the on-shell residues of the two quarks and the gluon,
we obtain, after UV renormalization,
\bea
 \left.\mathcal M^{(2)}_{\mathrm{fact}}\right|_{\mathrm{poles,\ incl.\ legs}}
 &=&\nn\\
 &&\hspace{-2.5cm}
 \left[\frac{g_s^2}{16\pi^2}\,\mathcal M^{(1)}
 \left\{
 (2C_F-C_A)
 \left(-\frac{1}{\epsilon^2}
       +\frac{1}{\epsilon}\ln\frac{-P^2-i0}{\mu^2}\right)
 -\frac{3C_F}{\epsilon}
 -\frac{5C_A-4T_Rn_f}{6\epsilon}
 \right\}\right]_{\mathrm{poles}} .
 \label{eq:check-topological}
\eea

Finally, consider $\mathcal C_{\mathrm{II}}^{\mathrm{soft}}$ in
Eq.~\eqref{eq:IR-II-soft-CT}. We retain both currents as written.
Expanding the two $G$ tensors cancels adjacent propagators through
$k\cdot(2p_j-k)=-(p_j-k)^2$, for $j=2,3$. The exact reduction is
\begin{align}
 \int\frac{d^d k}{(2\pi)^d}\mathcal C_{\mathrm{II}}^{\mathrm{soft}}(k)
 &=\frac{g_s^2 C_A}{32\pi^2}\mathcal M^{(1)}
 \big[4p_2\cdot p_3\,T(p_2,p_3)-B(p_2,p_3)\nonumber\\
 &\quad+B(p_3,q)-B(0,q)-4p_3\cdot q\,T(p_3,q)\nonumber\\
 &\quad+B(p_2,-q)-B(0,-q)+4p_2\cdot q\,T(p_2,-q)\nonumber\\
 &\quad+B(q,-q)-4q^2T(q,-q)\big].
\end{align}
The six bubble poles cancel. .
The change of variable $k\to P-k$ gives the same reduction for
$\mathcal C_{\mathrm I}^{\mathrm{soft}}$ with $p_2\to p_1$.
Hence
\begin{align}
 &\left[\int\frac{d^d k}{(2\pi)^d}
 (\mathcal C_{\mathrm I}^{\mathrm{soft}}+\mathcal C_{\mathrm{II}}^{\mathrm{soft}})
 \right]_{\rm poles}\nonumber\\
 &\quad=\left[\frac{g_s^2 C_A}{32\pi^2}\mathcal M^{(1)}
 \sum_{i=1}^2\left(4p_i\cdot p_3\,T(p_i,p_3)
 +4p_i\cdot q\,T(p_i,-q)
 -4p_3\cdot q\,T(p_3,q)+\frac2\epsilon\right)\right]_{\rm poles}.
 \label{eq:check-soft}
\end{align}
The three triangle double poles cancel within each counterterm.

On adding Eqs.~\eqref{eq:check-five}, \eqref{eq:check-J45}, and
\eqref{eq:check-soft}, the coefficients of $T(p_i,-q)$ and $T(p_3,q)$
vanish. The remaining explicit single pole from these three sources
is $-C_A/\epsilon$. Adding Eq.~\eqref{eq:check-topological}
and evaluating the massless triangles gives
\begin{align}
 \left.\mathcal M^{(2)}_{\rm ren}\right|_{\rm poles}
 &=\left[\frac{g_s^2}{16\pi^2}\mathcal M^{(1)}
 \bigg\{-\frac{2C_F+C_A}{\epsilon^2}
 +\frac{2C_F-C_A}{\epsilon}\ln\frac{-P^2-i0}{\mu^2}\right.\nonumber\\
 &\left.\qquad+\frac{C_A}{\epsilon}\sum_{i=1}^2
 \ln\frac{2p_i\cdot p_3}{\mu^2}
 -\frac{3C_F}{\epsilon}-\frac{11C_A-4T_Rn_f}{6\epsilon}
 \bigg\}\right]_{\rm poles}.
\end{align}
This agrees with the prediction of poles for this process made in \cite{Catani:1998bh} for $q\bar qg$ final state
continued to the annihilation channel. Since the tree amplitude
vanishes, the insertion multiplies the full $d$-dimensional
$\mathcal M^{(1)}$ of Eq.~\eqref{eq:IR-hard-one-loop-soft}.

\end{document}